\documentclass[showpacs,aps,prc,nofootinbib,showkeys,superscriptaddress,twocolumn]{revtex4-1}

\usepackage[colorlinks, urlcolor=blue,linkcolor=blue, anchorcolor=blue, citecolor=blue]{hyperref}
\usepackage[utf8]{inputenc}
\usepackage{natbib}
\usepackage{graphicx}
\usepackage{xcolor}
\usepackage{amsmath}
\usepackage{bm}
\usepackage[normalem]{ulem}

\newcommand{\V}{n}
\newcommand{\n}{{n}}
\newcommand{\ed}{{e}}

\newcommand{\peq}{p}

\newcommand{\beshydro}{{\sc BEShydro}}
\newcommand{\isd}{{\sc iS3D}}
\newcommand{\eq}{{\,=\,}}
\newcommand{\snn}{\sqrt{s_\mathrm{NN}}}
\newcommand{\etas}{$\eta_s$}
\newcommand{\tauI}{\tau_{i}}

\newcommand{\ld}[1]{{\color{black}#1}}
\newcommand{\xa}[1]{{\color{black}#1}}
\newcommand{\uh}[1]{{\color{black}#1}}

\graphicspath{{fig/}}

\begin{document}
\title{Baryon transport and the QCD critical point}

\author{Lipei Du}
\affiliation{Department of Physics, The Ohio State University, Columbus, Ohio 43210, USA}

\author{Xin An}
\affiliation{Department of Physics and Astronomy, University of North Carolina, Chapel Hill, North Carolina 27599, USA}

\author{Ulrich Heinz}
\affiliation{Department of Physics, The Ohio State University, Columbus, Ohio 43210, USA}

\date{\today}

\begin{abstract}
    
       Fireballs created in relativistic heavy-ion collisions at different beam energies have been argued to follow different trajectories in the QCD phase diagram in which the QCD critical point serves as a landmark. Using a (1+1)-dimensional model setting with transverse homogeneity, we study the complexities introduced by the fact that the evolution history of each fireball cannot be characterized by a single trajectory but rather covers an entire swath of the phase diagram, with the finally emitted hadron spectra integrating over contributions from many different trajectories. Studying the phase diagram trajectories of fluid cells at different space-time rapidities, we explore how baryon diffusion shuffles them around, and how they are affected by critical dynamics near the QCD critical point. We find a striking insensitivity of baryon diffusion to critical effects. Its origins are analyzed and possible implications discussed.
       
\end{abstract}

\maketitle

\section{Introduction}

One of the primary goals of nuclear physics \cite{Geesaman:2015fha} is studying the phase diagram of Quantum Chromodynamics (QCD), which is generally mapped to a plane expanded with temperature $T$ and baryon chemical potential $\mu$ axes \cite{Stephanov:2007fk}. First principles calculations from Lattice QCD show that, at zero $\mu$, the phase transition from a deconfined quark-gluon plasma (QGP) phase to a confined hadron resonance gas (HRG) phase from high to low temperature is a rapid but smooth crossover \cite{Aoki:2006we, Bazavov:2009zn, Borsanyi:2010cj, Bazavov:2011nk}. At large $\mu$, calculations of the phase transition using Lattice QCD are not yet available, since there the standard techniques suffer from the ``sign problem" \cite{PHILIPSEN201355,Ratti:2018ksb}. Nevertheless, theoretical models indicate that at large chemical potential, the phase transition is first order \cite{Stephanov:2007fk}, and this implies that a critical point exists at non-zero chemical potential \cite{Stephanov:1998dy,Stephanov:1999zu}, at the end of the first order phase transition line. Confirming the existence and finding the location of the hypothetical QCD critical point have attracted tremendous amount of attention over the last two decades \cite{Busza:2018rrf,Bzdak:2019pkr}.

Heavy-ion collisions are the main method to tackle these unsolved problems \cite{Berges:1998rc, Halasz:1998qr, Stephanov:1998dy, Stephanov:1999zu, Rajagopal:2000wf, Hatta:2003wn, Stephanov:2008qz, Fukushima:2010bq, Stephanov:2011pb, Luo:2017faz, Busza:2018rrf, Bzdak:2019pkr, Rajagopal:2019xwg, Du:2020bxp}. Such collisions have been carried out at different experimental facilities, such as the Large Hadron Collider (LHC) at CERN and the Relativistic Heavy-Ion Collider (RHIC) at Brookhaven National Laboratory, at various beam energies, and large sets of data have been accumulated. One of the most promising signatures of the QCD critical point is a non-monotonic beam energy dependence of higher-order cumulants of the fluctuations in the net proton production yields \cite{Stephanov:1998dy, Stephanov:1999zu, Hatta:2003wn, Stephanov:2008qz, Stephanov:2011pb, Luo:2017faz}. This is based on the idea that these observables are more sensitive to the correlation length of fluctuations of the chiral order parameter which, in the thermodynamic limit, diverges at the critical point \cite{RevModPhys.49.435}. Fireballs created in heavy-ion collisions at different beam energies should freeze out with correlation lengths that depend non-monotonically on the collision energy, and this should be reflected in the net baryon cumulants. Strongly motivated by this, a Beam Energy Scan (BES) program has been carried out at RHIC during the last decade. During a first campaign that ended in 2011 (BES-I), Au-Au collisions were studied at collision energies $\sqrt{s_\mathrm{NN}}$ from 200 GeV down to 7.7 GeV (BES-I) \cite{Adamczyk:2017iwn, Bzdak:2019pkr, Luo:2020pef}. A second campaign, BES-II, with significantly increased beam luminosity is expected to be completed this year, after having explored collision energies down to $\sqrt{s_\mathrm{NN}}=3.0$\,GeV in fixed-target mode. Additional experiments at even lower beam energies, probing the phase diagram in regions with even higher baryon chemical potential, are planned at the newly constructed FAIR and NICA facilities.

Unfortunately, the dynamical nature of the fireballs created in heavy-ion collisions renders attempts to confirm the above static equilibrium considerations experimentally anything but straightforward. Within their short lifetimes of several dozen yoctoseconds the fireballs' energy density decreases rapidly by collective expansion, from initially hundreds of GeV/fm$^3$ to well below 1~GeV/fm$^3$ at final freeze-out (see e.g. Refs.~\cite{Heinz:2013th,Busza:2018rrf}). The rapid dynamical evolution of the thermodynamic environment keeps the latter permanently out of thermal equilibrium such that critical fluctuations never reach their thermodynamic equilibrium distributions. In addition, in those parts of the fireball which pass through the quark-hadron phase transition close to the QCD critical point the dynamics of critical fluctuations is affected by ``critical slowing down'' \cite{Berdnikov:1999ph}. This is both a curse and a blessing: If critical fluctuations would relax quickly to thermal equilibrium, all memory of critical dynamics might have been erased from the hadronic freeze-out distributions by the time the hadron yields and momenta decouple. If, on the other hand, the dynamical evolution of fluctuations is slowed in the vicinity of the critical point, some signals of critical dynamics may survive until freeze-out but they will then most definitely not feature their equilibrium characteristics near the critical point \cite{Berdnikov:1999ph}.

Thus, to confirm or exclude the critical point via systematic model-data comparison, reliable dynamical simulations of off-equilibrium critical fluctuations and the associated final particle cumulants, on top of a well-constrained comprehensive dynamical description of the bulk medium at various beam energies, are indispensable \cite{Nahrgang:2011mv, Nahrgang:2011mg, Mukherjee:2015swa, Akamatsu:2016llw, Murase:2016rhl, Herold:2016uvv, Stephanov:2017ghc, Hirano:2018diu, Singh:2018dpk, Herold:2018ptm, Nahrgang:2018afz, Yin:2018ejt}. Recently, the Hydro+/++ framework \cite{Stephanov:2017ghc,An:2019hydro++} was developed for incorporating off-equilibrium fluctuations and critical slowing-down into hydrodynamic simulations, and some practical progress within simplified settings has since been made using this framework \cite{Rajagopal:2019xwg, Du:2020bxp}. On the other hand, while a fully developed and calibrated (2+1)-dimensional multi-stage description of heavy-ion collisions (including initial conditions + pre-hydrodynamic dynamics + viscous hydrodynamics + hadronic afterburner) exists (see, e.g., the most recent versions described in \cite{Nijs:2020ors, Nijs:2020roc, Everett:2020xug, Everett:2020yty}) and has met great phenomenological success at top RHIC and LHC energies \cite{Heinz:2013th, Busza:2018rrf, Bzdak:2019pkr}, such a comprehensive and fully validated framework is still missing for collisions at the lower BES energies. 

Compared to high-energy collisions at LHC and top RHIC energies, collisions at BES energies introduce a number of additional complications \cite{Shen:2020gef, Shen:2020mgh}. These include (i) a much more complex, intrinsically (3+1)-dimensional and temporally extended nuclear interpenetration stage and its associated dynamical  deposition of energy and baryon number \cite{Shen:2017bsr, Du:2018mpf, Martinez:2019jbu, Martinez:2019rlp, Shen:2017ruz}, (ii) the need to account for and properly propagate conserved charge currents for baryon number and strangeness \cite{Shen:2017ruz, Denicol:2018wdp, Monnai:2019hkn, Du:2019obx, Greif:2017byw, Fotakis:2019nbq, Shen:2020jwv}, (iii) a consistent treatment of singularities in the thermodynamic properties associated with the critical point \cite{RevModPhys.49.435, RevModPhys.50.85}, (iv) the aforementioned off-equilibrium nature of critical dynamics \cite{Stephanov:2017ghc, Nahrgang:2018afz, An:2019hydro++, Rajagopal:2019xwg, Du:2020bxp, Nahrgang:2020yxm}, \xa{and (v) the dynamical effects of nucleation and spinodal decomposition in the metastable and unstable regions associated with the first-order phase transition \cite{Randrup_2004,Sasaki_2007,Steinheimer_2012,An_2018}}. The situation is made even more complicated by the back-reaction of the non-equilibrium critical fluctuation dynamics on the bulk evolution of the medium. This back-reaction causes a potential dilemma: On the one hand, locating the critical point requires reliable calculations at various beam energies of critical fluctuations on top of a well-constrained bulk evolution of the fireball medium; on the other hand, the back-reaction of the off-equilibrium critical fluctuations from a critical point whose location is yet to be determined onto the medium evolution might interfere with the calibration of the latter and turn it into an impossibly complex iterative procedure whose convergence cannot be guaranteed.

Guidance on how (and perhaps even whether) to incorporate critical effects when constraining the bulk dynamics is direly needed, not least since dynamical simulations for low beam energies are computationally very expensive. Some effects on the bulk medium evolution arising from singularities in the thermodynamic properties of the QCD matter \cite{RevModPhys.49.435, RevModPhys.50.85}, by adding a critical point to the QCD Equation of State (EoS) \cite{Nonaka:2004pg, Parotto:2018pwx, Monnai:2019hkn, Monnai:2021kgu, Stafford:2021wik}  and/or explicitly including critical scaling of its transport coefficients, have been explored, with special attention to the bulk viscous pressure since critical fluctuations of the chiral order parameter, which couples to the baryon mass, can be directly related to a peak of the bulk viscosity near the critical point \cite{Karsch:2007jc, Moore:2008ws, NoronhaHostler:2008ju, Denicol:2009am}. The authors of \cite{Monnai:2016kud} showed that critical effects on the bulk viscous pressure have non-negligible phenomenological consequences for the rapidity distributions of hadronic particle yields, implying that critical effects might indeed play an important role in the calibration of the bulk medium.

In a similar spirit we study here critical effects on the bulk evolution in the baryon sector, by including the critical scaling of the relaxation time for the baryon diffusion current and of the baryon diffusion coefficient, as well as (in a simplified treatment) the critical contribution to the EoS \cite{Nonaka:2004pg, Parotto:2018pwx, Stafford:2021wik}. We study the phenomenological consequences of baryon diffusion in a system both away from and close to the QCD critical point; the former is essential for modeling heavy-ion collisions at the high end of the BES collision energy range \cite{Shen:2017ruz, Denicol:2018wdp, Monnai:2019hkn, Du:2019obx, Greif:2017byw, Fotakis:2019nbq, Shen:2020jwv}. Including only the baryon diffusion while neglecting other dissipative effects helps us to study its hydrodynamical consequences in isolation. A more comprehensive study including all dissipative effects simultaneously is left for future work. 

This paper is organized as follows. We discuss the hydrodynamic formalism with non-zero baryon diffusion current in Sec.~\ref{sec:hydro}. In Sec.~\ref{sec:critical}, we illustrate our setup near the QCD critical point, particularly by discussing the critical behavior of thermodynamic and transport coefficients for the hydrodynamic evolution. After completing the setup of the framework in Sec.~\ref{sec:setup} we discuss results for a fireball created at low beam energies in Sec.~\ref{sec:results}, with a focus on baryon diffusion current effects, first away from (Sec.~\ref{sec:diffnocp}) and then close to (Sec.~\ref{sec:diffcp}) the critical point, followed by a discussion of general features of the time-evolution of the diffusion current in Sec.~\ref{sec:tevol}. We summarize results and draw conclusions in Sec.~\ref{sec:summary}. In the Appendices, we discuss causality near the QCD critical point in App.~\ref{sec:causality}, estimate the size of the critical region in App.~\ref{sec:Delta_muT} and, finally,  validate the numerical methods used in this work in App.~\ref{sec:validation}.

Throughout this article we use natural units where factors of $\hbar,\,c$, and $k_B$ are not explicitly exhibited but implied by dimensional analysis. We also use Milne coordinates, $x^\mu=(\tau,x,y,\eta_s)$, where $\tau$ and $\eta_s$ are the (longitudinal) proper time and space-time rapidity, respectively, and are related to the Cartesian coordinates via $t{\,=\,}\tau\cosh\eta_s,\, z{\,=\,}\tau\sinh\eta_s$. We employ the mostly-minus convention for the metric tensor, $g^{\mu\nu}=\text{diag}\,(+1,-1,-1,-1/\tau^2)$.

\section{Viscous hydrodynamics with baryon diffusion}\label{sec:hydro}

In this section we discuss the viscous hydrodynamic framework including baryon diffusion. Hydrodynamics is an effective theory for describing long-wavelength degrees of freedom, which, from a macroscopic viewpoint, can be formulated by conservation laws of hydrodynamic variables that are ensemble-averaged at certain coarse-grained scales \cite{Heinz:2013th, Jeon:2015dfa}. In heavy-ion collisions, the conserved quantities are energy, momentum, and various charges, including net baryon charge, electric charge and strangeness. In this work, we only study the net baryon charge; the extension to incorporate other conserved charges is conceptually straightforward \cite{Fotakis:2019nbq} and will be studied elsewhere. The conservation equations for energy-momentum and net baryon charge are formulated covariantly as
\begin{subequations}\label{eq:conservation}
\begin{align}
d_\mu T^{\mu\nu} &= 0\,,\label{eq:conservation1}\\
d_\mu N^\mu &= 0\,,\label{eq:conservation2}
\end{align}
\end{subequations}
where $d_\mu$ is the covariant derivative in an arbitrary coordinate system (Milne coordinates here), $T^{\mu\nu}$ and $N^\mu$ are the energy-momentum tensor and (net) baryon current, respectively. In a given arbitrary reference frame, they can be decomposed into the ideal and dissipative parts as
\begin{subequations}\label{eq:constitutive_decomposition}
\begin{align}
T^{\mu\nu} &= T^{\mu\nu}_\mathrm{id}+\Pi^{\mu\nu}\;,\\
N^{\mu} &= N^\mu_\mathrm{id} + n^{\mu}\,,
\end{align}
\end{subequations}
where $T^{\mu\nu}_\mathrm{id}$ and $N^\mu_\mathrm{id}$ are the ideal parts which are well defined in local equilibrium, while $\Pi^{\mu\nu}$ and $n^{\mu}$ are the dissipative components describing the deviations from local equilibrium. The former can be expressed as
\begin{subequations}\label{eq:constitutive_ideal}
\begin{align}
T^{\mu\nu}_\mathrm{id} &= \ed u^{\mu}u^{\nu}-\peq\Delta^{\mu\nu}\;,\label{eq:constitutive_ideal1}\\
N^{\mu}_\mathrm{id} &= n u^{\mu}\,,\label{eq:constitutive_ideal2}
\end{align}
\end{subequations}
where $\ed$ and $n$ are the energy density and baryon density in the local rest frame, $u^\mu$ is the four-velocity of the fluid element (normalized as $u^2=1$), $p$ the pressure given by the EoS, $p=p(e, n)$, and $\Delta^{\mu\nu} \equiv g^{\mu\nu} - u^{\mu}u^{\nu}$. The local rest frame in this work is chosen as the Landau frame specified by the Landau matching conditions \cite{landau2013statistical}
\begin{equation}\label{eq:Landau-matching}
    u_\mu T^{\mu\nu}=\ed u^\nu\,, \quad u_\mu N^\mu=n\,, 
\end{equation}
which implies $u_\mu n^{\mu} = 0$ and $u_\mu\Pi^{\mu\nu}=0$.

The dissipative term $\Pi^{\mu\nu}$ in the energy-momentum tensor can be written as $\Pi^{\mu\nu}=-\Pi\Delta^{\mu\nu}+\pi^{\mu\nu}$, where $\Pi$ is the bulk viscous pressure, and $\pi^{\mu\nu}$ the shear stress tensor. The dissipative term $n^\mu$ in the net baryon current describes its non-zero spatial components in the local rest frame of the fluid. The evolution of these dissipative terms are governed by both microscopic and macroscopic physics, and thus their equations of motion can not be obtained directly from conservation laws. We use the evolution equations from the Denicol-Niemi-Molnar-Rischke (DNMR) theory \cite{Denicol:2010xn,Denicol:2012cn}, which uses the methods of moments of the Boltzmann equation. In this work, to isolate the effects from baryon diffusion current $n^\mu$ clearly, we shall ignore the dissipative effects from $\pi^{\mu\nu}$ and $\Pi$, focusing only on $n^\mu$.

The equation of motion for $n^\mu$ from the DNMR theory is an Israel-Stewart type equation,
\begin{equation}\label{eq:IS_nmu}
\tau_{n}\dot{n}^{\left\langle \mu \right\rangle}+n^{\mu } = \kappa_n
 \nabla^{\mu}\alpha+ {\cal J}^\mu\,,
\end{equation}
where $\dot{n}^{\left\langle \mu \right\rangle} \equiv \Delta^\mu_\nu\dot{n}^\nu$ (the overdot denotes the covariant time derivative $D \equiv u_\mu d^{\mu}$), $\kappa_n$ is the baryon diffusion coefficient (also referred to as the baryon conductivity), $\alpha\equiv\mu/T$ is the chemical potential $\mu$ in the unit of temperature $T$,  and $\tau_{n}$ is the relaxation time, on whose scale baryon current relaxes towards its Navier-Stokes limit:
\begin{equation}
    n^{\mu}_{\rm NS}\equiv\kappa_n \nabla^{\mu}\alpha\,;
\end{equation}
here $\nabla^\mu \equiv \Delta^{\mu\nu}d_\nu$ is the spatial gradient in the local rest frame. The term ${\cal J}^\mu$ contains higher order gradient contributions \cite{Denicol:2010xn,Denicol:2012cn}. Rewriting Eq.~\eqref{eq:IS_nmu} as a relaxation equation,
\begin{equation}
\label{eq:IS_nmu2}
    \dot{n}^{\left\langle \mu \right\rangle} = -\frac{1}{\tau_{n}}(n^{\mu }-\kappa_n \nabla^{\mu}\alpha) + \frac{1}{\tau_{n}}{\cal J}^\mu\,,
\end{equation}
shows that $\nabla^{\mu}\alpha$ is the driving force for baryon diffusion while $\kappa_n$ controls the strength of the baryon diffusion flux arising in response to this force. $\tau_n$ characterizes the response time scale. We note that both $\tau_{n}$ and $\kappa_n$ depend on the microscopic properties of the medium, which have been calculated in various theoretical frameworks, including kinetic theory \cite{Denicol:2018wdp} and holographic models \cite{Son:2006em,Rougemont:2015ona}. In principle, they can also be constrained phenomenologically by data-driven model inference, but as of today such studies are still very limited for baryon evolution. 

Rewriting Eq.~\eqref{eq:IS_nmu} more specifically for baryon diffusion, we arrive at
\begin{equation}
\label{eq:IS_nmu3}
    u^\nu\partial_\nu \V^\mu = \frac{\kappa_n}{\tau_\V} \nabla^{\mu}\alpha - \frac{\V^{\mu}}{\tau_{\V}}  -\frac{\delta_{nn}}{\tau_n}n^\mu\theta - n^\nu u^\mu D u_\nu
    - u^\alpha \Gamma^\mu_{\alpha\beta} n^\beta\,,
\end{equation}
where $\theta\equiv d\cdot u$, the $n^\mu\theta$-term arises from ${\cal J}^\mu$ (the only term we keep from ${\cal J}^\mu$ as given in \cite{Du:2019obx}), $\delta_{nn}$ is the associated transport coefficient, and the last two terms come from rewriting $\dot{n}^{\left\langle \mu \right\rangle}$ explicitly, with $\Gamma^\mu_{\alpha\beta}$ being the Christoffel symbols. Eq.~(\ref{eq:IS_nmu3}) is the equation we use to evolve the baryon diffusion current in this work. We remark that the Navier-Stokes limit of the baryon diffusion current
can be rewritten in terms of density and temperature gradients,
\begin{subequations}
\label{eq:nmu_NS_decomposition}
\begin{eqnarray}
n^{\mu}_{\rm NS} 
= D_B\nabla^{\mu}n+D_T\nabla^{\mu}T\,,
\end{eqnarray}
where the two coefficients are
\begin{equation}
\label{eq:D_BT}
    D_B=\frac{\kappa_n}{T\chi}, \quad D_T=\frac{\kappa_n}{Tn}\left[\left(\frac{\partial p}{\partial T}\right)_n-\frac{w}{T}\right]\,.
\end{equation}
\end{subequations}
Here $\chi\equiv(\partial n/\partial\mu)_T$ is the isothermal susceptibility, and $w=\ed+p$ is the enthalpy density. We note that the gradient expansion is not unique, and writing it in different ways can be used to explore individual contributions separately (see App.~\ref{sec:gradient_validation}). We will discuss the benefits of decomposing the Navier-Stokes limit as Eq.~(\ref{eq:nmu_NS_decomposition}) where critical singularities manifest themselves in Sec.~\ref{sec:critical}.


For later convenience of discussing critical behavior, we also introduce the heat diffusion coefficient,
\begin{equation}\label{eq:D_p}
    D_p=\frac{\lambda_T}{c_p}\,,
\end{equation}
where $c_p\equiv nT(\partial m/\partial T)_p$ is the specific heat, with $m\equiv s/n$, i.e., the entropy per baryon density \cite{Son:2004iv, Stephanov:2017ghc}; $\lambda_T$ is the thermal conductivity, which can be related to the baryon diffusion coefficient $\kappa_n$ by
\begin{equation}\label{eq:kappa-lambda}
    \lambda_T = \left(\frac{w}{nT}\right)^2\kappa_n\,.
\end{equation}
Using Eq.~\eqref{eq:kappa-lambda} one can also relate the heat diffusion coefficient $D_p$ to $D_B$ by
\begin{equation}\label{eq:Dp-DB}
    D_p=\left(\frac{w^2\chi}{n^2Tc_p}\right)D_B.
\end{equation}

The above system of hydrodynamic equations is closed by the EoS, either in the form $p(\ed, n)$ or, equivalently, through the pair of relations $\bigl(\mu(\ed, n),\,T(\ed, n)\bigr)$. Again, the EoS is controlled by microscopic physics. We discuss the EoS used in this study in Sec. \ref{sec:setup} below.

\section{Critical behavior}
\label{sec:critical}

In this section we focus on the effects of the QCD critical point on baryon transport in a relativistic QCD fluid. Critical phenomena, albeit ubiquitous, can be classified into certain universality classes determined by the effective degrees of freedom and symmetries of the system. It has been well argued that the QCD critical point belongs to the static universality class of the 3-dimensional Ising model \cite{Berges:1998rc, Halasz:1998qr}, and the dynamical universality class of Model H in the Hohenberg-Halperin classification \cite{RevModPhys.49.435, Son:2004iv}. It is also believed that the critical point, if it exists, is beyond the reach of first-principles approaches such as Lattice QCD, and consequently little else is known beyond its universality classification. 

Much work has been done to lift the fog from the physics \xa{lurking} beyond the above universality argument. First, a robust construction of a family of EoS exhibiting the appropriate universal critical properties and matching to existing Lattice QCD calculations has been proposed \cite{Parotto:2018pwx, Stafford:2021wik}. The EoS encodes the microscopic properties of the QCD matter and is indispensable for solving the macroscopic hydrodynamic equations. Second, a hydrodynamic framework incorporating fluctuations and critical slowing-down has been established, in order to overcome the breakdown of hydrodynamics near the critical point. A deterministic framework, known as hydro-kinetic theory, extends conventional hydrodynamics by consistently including fluctuations as additional dynamic degrees of freedom (modes) \cite{Akamatsu:2018vjr, Martinez:2018, An:2019osr, An:2020vri}. The feedback of fluctuating modes renormalizes the bare hydrodynamic variables and gives rise to a delayed response in the form of so-called ``long-time tails''. The hydro-kinetic approach can be implemented in the critical regime where the fluctuating modes relax to equilibrium on parametrically long time scales \cite{Stephanov:2017ghc, An:2019hydro++} (see also the reviews \cite{Shen:2020mgh, An:2020jjk}). 

Of course, the inclusion of fluctuations does not by itself cure pathological issues such as acausality or instability of the underlying hydrodynamic framework that arise when straightforwardly extending the non-relativistic Navier-Stokes equations into the relativistic domain \cite{Hiscock:1985zz, Hiscock:1987zz}. The most-widely used resolution of these issues follows the approach pioneered in Ref.~\cite{Israel:1979}, by elevating the dissipative components of the energy-momentum tensor to dynamical degrees of freedom subject to their own relaxation-type equations (for which Eq.~\eqref{eq:IS_nmu2} for the baryon diffusion current is an example). In this approach the causality and stability conditions can be shown to continue to hold in the proximity of the critical point (see App.~\ref{sec:causality}).

In the following subsections we discuss how we implement the static and dynamic universal critical behavior, using a simplified setup. Possible future improvements using a more realistic implementation will be discussed in the conclusions (Sec.~\ref{sec:summary}).

\subsection{Implementation of static critical behavior}

\begin{figure}[!tbp]
\begin{center}
    \includegraphics[width= 0.40\textwidth]{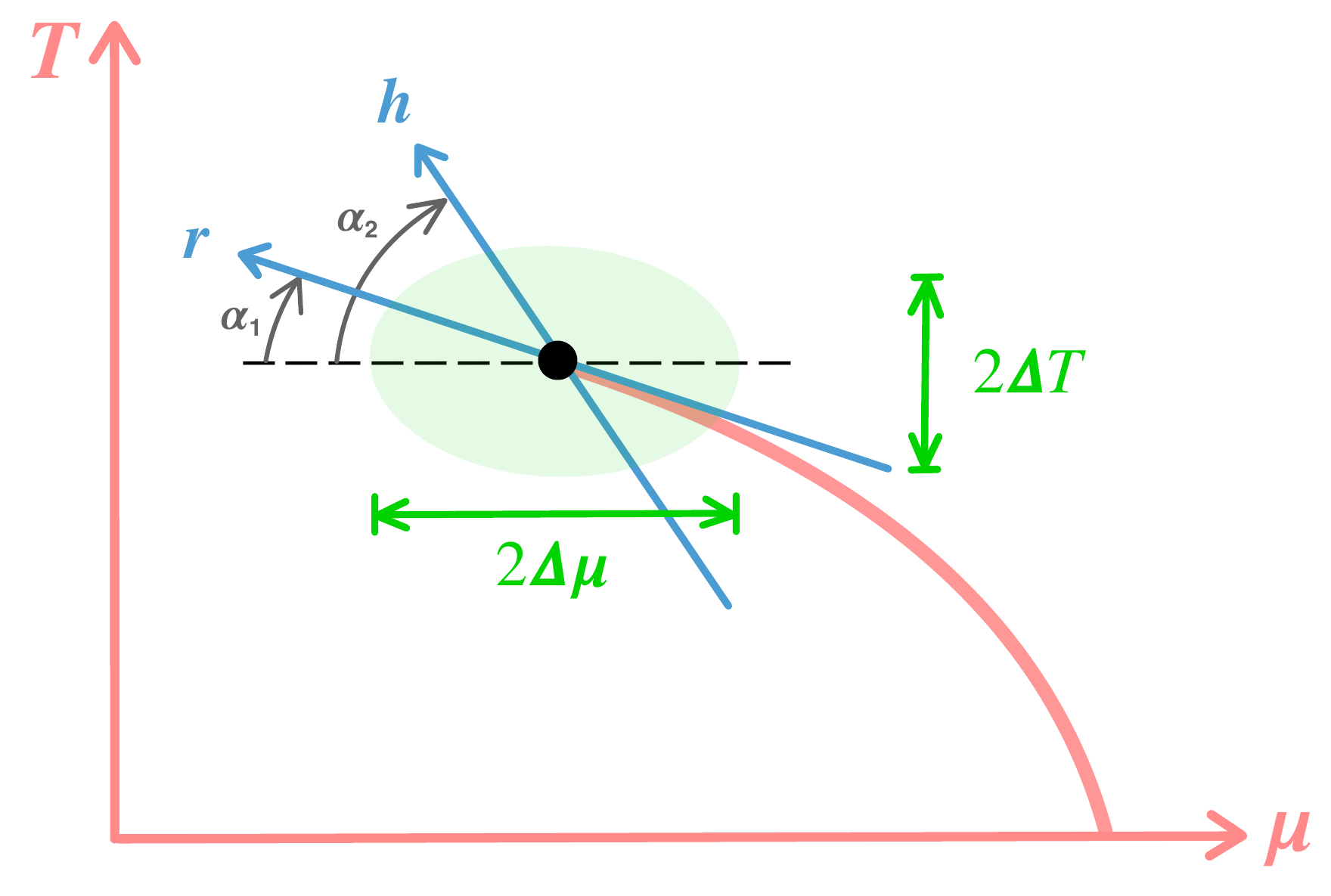}
    \caption{A schematic phase diagram for the mapping from Ising variables $(r, h)$ to the QCD phase diagram coordinate $(\mu,T)$. The black circle ($\mu_c, T_c$) the is QCD critical point, in the critical region with width $(2\Delta\mu, 2\Delta T)$.}
    \label{fig:phase_diagram}
\end{center}
\end{figure}

\begin{figure}[!tbp]
\begin{center}
    \includegraphics[width= 0.45\textwidth]{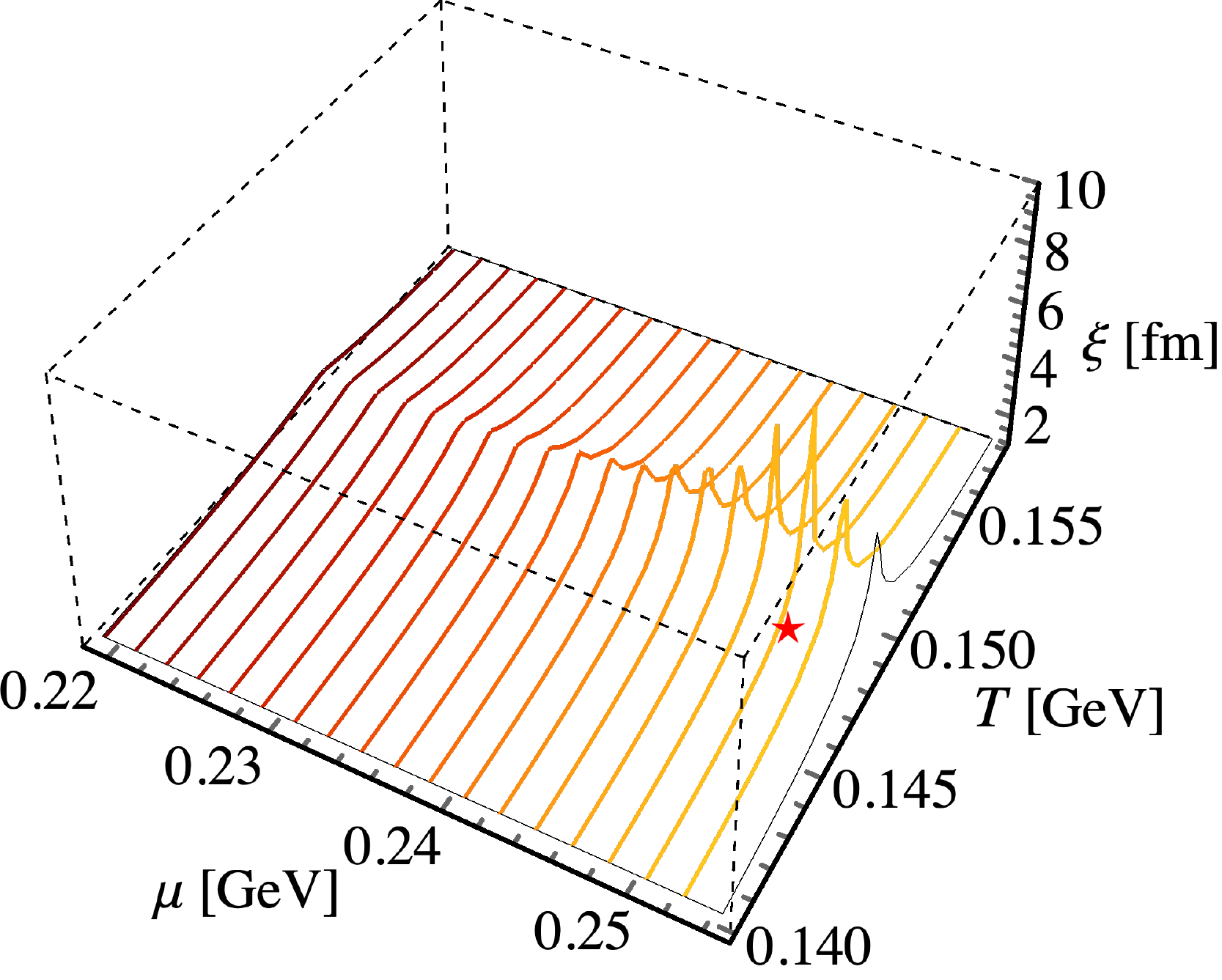}
\end{center}
\caption{%
    Distribution of the correlation length $\xi(\mu,T)$, with a critical point located in the $T$-$\mu$ plane at $T_c{\,=\,}0.149$\,GeV and ${\mu}_c{\,=\,}0.25$\,GeV (indicated by the red star), parametrized by Eq.~\eqref{eq:xi(mu,T)} with the parameters in Eq.~\eqref{eq:non-universal-parameters}.}
\label{fig:correlation_length}
\end{figure}

One significant feature of critical phenomena is that, when the system approaches a critical point adiabatically, the {\em equilibrium} correlation length, which is typically microscopically small, becomes macroscopically large and eventually diverges. With the purpose of identifying qualitative signatures of a critical point, we characterize all equilibrium quantities exhibiting critical behavior in terms of their parametric dependence on the correlation length.\footnote{%
    If we knew the critical EoS explicitly, these dependencies would naturally follow from the thermodynamic identities relating these quantities to the thermal equilibrium partition function.}
We parametrize the correlation length as follows:
\begin{eqnarray}\label{eq:xi(mu,T)}
    \xi(\mu,T) =&\,\xi_0(\mu,T)\left\{\tanh [f(\mu,T)]\left(1-\left(\frac{\xi_0}{\xi_\mathrm{max}}\right)^{\frac{2}{\nu}}\right)\right.\nonumber\\
    &\left.+\left(\frac{\xi_0}{\xi_\mathrm{max}}\right)^{\frac{2}{\nu}}\right\}^{-\frac{\nu}{2}}\,.
\end{eqnarray}
Here $\xi_0(\mu,T)$ is the non-critical correlation length (measured far away from the critical point) while $\xi_\mathrm{max}$ is an infrared cutoff regulating the divergence at the critical point by implementing a maximum value for the correlation length. The crossover between the critical and non-critical regimes is characterized by the hyperbolic function $\tanh[f(\mu,T)]$, where
\begin{eqnarray}\label{eq:f(mu,T)}
    f(\mu,T)&=&\left|\frac{(\mu{-}\mu_c)\cos{\alpha_1}-(T{-}T_c)\sin{\alpha_1}}{\Delta \mu}\right|^2
\nonumber\\
    &+&\left|\frac{(\mu{-}\mu_c)\sin{\alpha_1}+(T{-}T_c)\cos{\alpha_1}}{\Delta T}\right|^{\frac{2}{\beta\delta}}.
\end{eqnarray}
In the above expression $(T_c,\mu_c)$ is the location of critical point, and $\Delta \mu$ and $\Delta T$ characterize the extent of the critical region along the $\mu$ and $T$ axes of the phase diagram; $\alpha_1$ is the angle between the crossover line ($h=0$ axis in the Ising model) and the negative $\mu$ axis (see Fig.~\ref{fig:phase_diagram}); $\nu=2/3$, $\beta=1/3$ and $\delta=5$ approximate the critical exponents of the 3-dimensional Ising universality class \cite{Berges:1998rc, Halasz:1998qr}. Eq.~\eqref{eq:xi(mu,T)} is designed to ensure the following properties:
\begin{itemize}
\setlength{\itemindent}{-1.3em}
    \item[] 1) $\xi=\xi_\mathrm{max}$ when $\mu=\mu_c$ and $T=T_c$\,;
    \item[] 2) $\xi\sim|T-T_c|^{-\frac{\nu}{\beta\delta}}$ when $\mu=\mu_c$ and $|T- T_c|\lesssim\Delta T$\,;
    \item[] 3) $\xi\sim|\mu-\mu_c|^{-\nu}$ when $T=T_c$ and $|\mu-\mu_c|\lesssim\Delta\mu$\,;
    \item[] 4) $\xi=\xi_0$ when $|\mu-\mu_c|\gg\Delta\mu$ and/or $|T-T_c|\gg\Delta T$\,.
\end{itemize}
%
%
To limit the number of free parameters, we ignore the $T$ and $\mu$ dependence of the non-critical correlation length $\xi_0$ and parametrize the crossover line as \cite{Bellwied:2015rza}
\begin{equation}
    \frac{T(\mu)}{T_0}=1-\kappa_2\left(\frac{\mu}{T_0}\right)^2+\mathcal{O}(\mu^4)\,,
\end{equation}
where $T_0=155$ MeV and $\kappa_2=0.0149$ are the transition temperature and the curvature of the transition line $T(\mu)$ at $\mu=0$. Location of the critical point $(T_c,\mu_c)$ is assumed to be on the crossover line \cite{Parotto:2018pwx}, and as a consequence
\begin{equation}
    \alpha_1=\arctan\left(\frac{2\kappa_2\mu_c}{T_0}\right).
\end{equation}
Thus $T_c$ and $\alpha_1$ are determined once $\mu_c$ is provided. Based on the above discussion we choose the following parameter values:
\begin{gather}
\xi_0 = 1\,\textrm{fm},\quad \xi_\mathrm{max} = 10\,\textrm{fm}, \nonumber\\
\mu_c=250\,\textrm{MeV}, \quad T_c = 149\,\textrm{MeV}, \quad \alpha_1=4.6^\circ\,, \label{eq:non-universal-parameters}\\
\Delta\mu = 92\,\textrm{MeV}, \quad \Delta T=18\,\textrm{MeV}. \nonumber
\end{gather}
Among these, $\Delta\mu$ and $\Delta T$ are determined by additional parameters provided in Eq.~\eqref{eq:setup_Delta_muT} of App.~\ref{sec:Delta_muT}. With those parameters, we visualize the correlation length as function of $(\mu,T)$ in Fig.~\ref{fig:correlation_length}.

Several comments are in order: First, our parametrization of the correlation length applies to the crossover region in the left part of the QCD $T$-$\mu$ phase diagram, at $\mu<\mu_c$, and not to the presumed first-order phase transition at $\mu>\mu_c$ where the theoretical description is complicated by possible phase coexistence and metastability \cite{An_2018}. This suggests choosing the collision beam energy sufficiently high to avoid the latter situation, but not too high to be far from the critical point. Motivated by experimental hints \cite{Adam:2020unf} and earlier theoretical studies \cite{Monnai:2016kud,Denicol:2018wdp}
we here set $\snn\eq19.6\,$GeV. Second, although the correlation length diverges in the thermodynamic limit, heavy-ion collisions create small, rapidly expanding QGP droplets in which finite-size and finite-time effects as well as the critical slowing down \cite{Stephanov:1999zu,Berdnikov:1999ph} prevent the correlation length from growing to infinity. A robust estimate for the largest correlation length the system might achieve in this dynamical environment is about 3\,fm \cite{Berdnikov:1999ph}. The system will thus never get close to our static infrared cutoff $\xi_{\rm max}=10$\,fm, and our final predictions turn out not to be sensitive to the precise value of this cutoff. 

Once all thermodynamic quantities and transport coefficients (introduced in the following subsection) are parametrized in terms of $\xi$ as given in Eq.~\eqref{eq:xi(mu,T)}, they are defined in both the non-critical and critical regions and thus ready for use in dynamical simulations describing the trajectory of the QGP fireball through the phase diagram. For economy we include in the following discussion not only the dynamic (transport) coefficients but also the thermal susceptibility $\chi$ and the specific heat $c_p$ which are static (thermodynamic) coefficients.

\subsection{Implementation of dynamic critical behavior}
\label{sec:critical_scaling}

Near the critical point, fluctuations at the length scale $\xi$ significantly modify the physical transport coefficients, giving rise to their correlation length dependence. In Model H, the shear stress tensor and bulk viscous pressure play important roles in critical dynamics through fluid advection \cite{Rajagopal:2019xwg, Du:2020bxp}. We shall only focus on critical dynamics arising from fluctuations in the hydrodynamic regime (i.e., carrying small wavenumbers/frequencies, $q\lesssim1/\xi$), as large-wavenumber fluctuations equilibrate fast compared to the hydrodynamic evolution rate $\omega\sim c_sk$, with $c_s$ being the speed of sound. Feedback from off-equilibrium fluctuations that are non-analytic in $\omega$ or $k$, commonly referred to as long-time tails, is suppressed by phase space \cite{Rajagopal:2019xwg, Du:2020bxp} and will be neglected. In other words, the scaling parametrizations below derive from equilibrium fluctuations for thermodynamic quantities and from analytic non-equilibrium fluctuations for  transport coefficients.

In the presence of a bulk viscous pressure, the critical contribution to bulk viscosity diverges as $\xi^z$ where $z=3$ for the QCD critical point \cite{RevModPhys.49.435}. Besides, the associated relaxation time for the bulk viscous pressure also diverges as $\xi^3$ \cite{Monnai:2016kud}. In this case, the relaxation rate of the fluctuation modes contributing to bulk viscous pressure is much smaller than the typical hydrodynamic frequency, and they can no longer be treated hydrodynamically, requiring instead an extended framework such as Hydro+ \cite{Stephanov:2017ghc,An:2019hydro++}. In this work we focus on effects from baryon diffusion and neglect bulk and shear stress effects on the bulk evolution dynamics of the fireball; however, we use the same critical scaling laws for the transport coefficients in Model H as if they were restored (i.e., with non-vanishing shear viscosity $\eta$ and bulk viscosity $\zeta$).

The following second-order thermodynamic coefficients (isothermal susceptibility $\chi$ and specific heat $c_p$) as well as the first-order transport coefficients (baryon diffusion coefficient $\kappa_n$ and thermal conductivity $\lambda_T$) scale with the correlation length as \cite{RevModPhys.49.435}
    \begin{equation}\label{eq:coeff_scaling}
        \chi\sim c_p\sim \xi^2\,, \quad \kappa_n\sim\lambda_T\sim \xi\,,
    \end{equation}
where the exponents are rounded to their nearest integers for simplicity. Therefore, according to Eqs.~\eqref{eq:D_BT} and \eqref{eq:D_p},
\begin{equation}
\label{eq:D_scaling}
    D_T\sim \xi, \quad D_B\sim D_p\sim \xi^{-1}. 
\end{equation}
In this work we apply the following parametrizations:
\begin{equation}\label{eq:chi_kappa_scaling}
    \chi=\chi_0\left(\frac{\xi}{\xi_0}\right)^2, \quad \kappa_n=\kappa_{n,0}\left(\frac{\xi}{\xi_0}\right),
\end{equation}
where $\xi_0$ is the non-critical correlation length, $\kappa_{n,0}$ is the non-critical value of baryon diffusion coefficient (see Eq.~\eqref{eq:kappa_kinetic} below), $\chi_0$ is the isothermal susceptibility evaluated in the non-critical region, $\chi_0\equiv(\partial n_0/\partial\mu)_T$ where $n_0$ is the non-critical baryon density.\footnote{%
    While the notational distinction between $\chi$ and $\chi_0$ is needed here for clarity, we generally drop the subscript ``0'' for thermodynamic quantities away from the critical region elsewhere to avoid clutter.} 
With Eq.~\eqref{eq:chi_kappa_scaling} the parametrizations with critical scaling for $D_B$, $D_T$ and $\lambda_T$ are readily obtained from Eqs.~\eqref{eq:D_BT} and \eqref{eq:kappa-lambda}.

We now turn to the critical behavior of the relaxation time $\tau_n$. It is worth remembering that the Israel-Stewart type equations (cf. Eq.~\eqref{eq:IS_nmu}) provide an ultraviolet completion of the naive (Landau-Lifshitz) hydrodynamic theory. The microscopic relaxation times associated with the new dissipative dynamical degrees of freedom (such as, in our case here, the baryon diffusion current $n^\mu$) play the role of ultraviolet regulators which modify the short-distance (high-frequency) behavior of the theory. For the baryon diffusion current $n^\mu$, $\tau_n$ characterizes the relaxation time to its Navier-Stokes limit $n^\mu_{\text{NS}}$ (which is zero in a homogeneous background). Since $n^\mu$ can only equilibrate as long as {\it all} fluctuating degrees of freedom contributing to $n^\mu$ also equilibrate, $\tau_n$ can be considered as the typical equilibration time scale of the slowest fluctuation mode near the critical point. Indeed, in Hydro+/++, the non-hydrodynamic slow-mode evolution equations for critical fluctuations with typical momenta $q\sim\xi^{-1}$ have the similar structure as the Israel-Stewart relaxation equations for the dissipative flows arising from thermal fluctuations with wavenumbers $\xi^{-1}\sim T$ (see Sec.~\ref{sec:hydro+} for detailed discussion). Here we assume the scale hierarchy adopted in \cite{An:2019hydro++}, i.e., the fluctuation wavenumber $q$ is much bigger than the gradient wavenumber $k$, but still much small than the inverse of thermal length $T$, i.e., $k\ll q\ll T$. As already mentioned, Israel-Stewart type equations neglect the non-analytic contributions from long-time tails which we argued above to be negligible (see also Sec.~\ref{sec:hydro+}).

According to Ref.~\cite{An:2019hydro++}, the {\it slowest} mode contributing to $n^\mu$ is the diffusive-shear correlator between the entropy per baryon density fluctuations $\delta m\equiv\delta (s/n)$ and the flow fluctuations $\delta u_\mu$, i.e., $G_{m\mu}\sim\langle\delta m \delta u_\mu \rangle$. The relaxation rate for this mode with wavenumber $q$ is given by $\Gamma_G(q)=(\gamma_\eta+D_p)q^2$ where $\gamma_\eta=\eta/w$. The two contributions to this rate stem from the relaxation of the shear stress and of the baryon diffusion, respectively.
Near the critical point $\Gamma_G$ is dominated by  contributions with typical wavenumbers $q\sim 1/\xi$. Given $D_p\sim\xi^{-1}$ and approximately $\gamma_\eta\sim\eta\sim\xi^0$ \cite{RevModPhys.49.435}, one finds $\Gamma_G(q)=(\gamma_\eta+D_p)q^2|_{q\sim\xi^{-1}}\sim\xi^{-2}$ and hence $\tau_G = \Gamma_G^{-1} \sim \xi^2$.\footnote{%
     Another mode contributing to the baryon diffusion current is the pressure-shear mode $G_{p\mu}\sim\langle\delta p\delta u_\mu\rangle$ \cite{An:2019hydro++}. Its relaxation rate at wavenumber $q$ is $(\gamma_\zeta+\frac{4}{3}\gamma_\eta+\gamma_p)q^2$ where  $\gamma_\zeta=\zeta/w$, $\gamma_\eta=\eta/w$, and $\gamma_p=\kappa_n c_s^2Tw(\partial\alpha/\partial p)_m^2$. In the presence of bulk viscosity (as we assume in order to ensure the correct scaling in Model H), the relaxation rate for $G_{p\mu}$ is dominated by $\gamma_\zeta q^2|_{q\sim\xi^{-1}}\sim\xi$ considering $\gamma_\zeta\sim\zeta\sim\xi^3$, which is much faster than the rate for the diffusive-shear mode which scales like $\Gamma_G\sim\xi^{-2}$. Even in the absence of the viscosities (i.e. for $\eta=\zeta=0$), the dominated relaxation rate $\gamma_p q^2|_{q\sim\xi^{-1}} \sim \xi^{-1}$ is still faster than $\Gamma_G \sim \xi^{-2}$. Therefore, the contribution from the pressure-shear mode, which is not the slowest, can be neglected.}
Thus it is natural to expect $\tau_n\sim\tau_G\sim\xi^2$. We therefore parametrize $\tau_n$ as
\begin{equation}\label{eq:tau_n_xi}
    \tau_n=\tau_{n,0}\left(\frac{\xi}{\xi_0}\right)^2,
\end{equation}
where $\tau_{n,0}$ is the non-critical relaxation time (given explicitly below in Eq.~\eqref{eq:taun_kinetic}). As we shall discuss in App.~\ref{sec:causality}, the parametrization \eqref{eq:tau_n_xi} ensures causality.
 
As an aside, let us comment on the consequences, had we tried to ensure the absence of shear stress by demanding that $\eta=0$. In this case $\Gamma_G(q) = D_p q^2|_{q\sim\xi^{-1}}\sim\xi^{-3}$, and therefore $\tau_n=\Gamma_n^{-1}\sim\tau_G\sim\xi^3$, which is larger compared to that in the case with shear stress. This arises from the fact that, near the critical point, the shear mode ($\delta u_\mu$) relaxes to equilibrium parametrically faster than the diffusive mode ($\delta m)$. As a result, its dissipation changes the scaling exponent of the relaxation time of the diffusive-shear-mode correlator.

Substituting Eq.~\eqref{eq:xi(mu,T)} into Eqs.~\eqref{eq:chi_kappa_scaling} and \eqref{eq:tau_n_xi} we arrive at a complete set of relevant thermodynamic quantities and transport coefficients (i.e., $\chi, \kappa_n$ and $\tau_n$) as explicit functions of $T$ and $\mu$ that hold in the entire crossover domain of the QCD phase diagram, both far away from and within the critical region.

\subsection{Connection to Hydro+}
\label{sec:hydro+}

We close this section by discussing the connection between the Israel-Stewart-like second-order hydrodynamic equations (cf. Eq.~\eqref{eq:IS_nmu2}) and the evolution equations proposed in the Hydro+ formalism \cite{Stephanov:2017ghc}.

\ld{When comparing the conventional Israel-Stewart formalism without critical effects with the Hydro+ framework, we note that} both approaches add {\em non-hydrodynamic} degrees of freedom to the conventional hydrodynamic framework, together with their associated relaxation time scales that are not negligible in rapidly evolving systems. The Hydro+ formalism distinguishes itself by the fact that the dynamics of some of these additional non-hydrodynamic modes, the critical slow modes, are controlled by a separate small parameter (different from the \xa{conventional} Knudsen number for the hydrodynamic gradient expansion) that characterizes the relative ``slowness'' of the relaxation of the critical slow modes when compared with standard dissipative effects arising from thermal fluctuations.

In our study \ld{the critical scaling for the transport coefficients is included in the Israel-Stewart formalism, and consequently it becomes directly comparable to the Hydro+ framework with a single wavenumber ($q\sim\xi^{-1}$) slow mode. To derive the former from the latter,} we introduce a non-hydrodynamic slow mode described by a vector field, in contrast to the scalar field considered \xa{as a primary example} in \ld{the Hydro+ formalism} \cite{Stephanov:2017ghc}.\footnote{%
    If, however, a scalar slow mode does exist (such as the bulk viscous pressure $\Pi$ in the critical regime \cite{Monnai:2016kud}), its dynamics can be formulated similarly, see Ref.~\cite{Stephanov:2017ghc}.} 
We denote this field by $\phi^\mu$ and demand that it is a transverse vector, i.e., $u\cdot\phi=0$, for the sake of later convenience. \ld{This vector field will be treated as the slowest mode contributing to $n^\mu$, corresponding to $G_{m\mu}\sim\langle\delta m \delta u_\mu \rangle$ discussed above.} 

Including this field the first law of thermodynamics is generalized as follows:
\begin{equation}\label{eq:gener_1st}
    ds_{(+)}=\beta_{(+)}de-\alpha_{(+)}dn+\pi\cdot d\phi\,.
\end{equation}
Here and below the subscript $(+)$ labels the generalized quantities by taking into account the additional contributions arising from the field $\phi^\mu$ (more precisely, from its deviation from its equilibrium value $\bar\phi^\mu$). The variable $\pi^\mu$ is thermodynamically conjugate to $\phi^\mu$, playing the role of a generalized thermodynamic potential with the constraint $u\cdot\pi=0$. $\beta_{(+)}$ and $\alpha_{(+)}$ are the associated inverse temperature and chemical potential in units of the temperature, respectively. We require that in thermal equilibrium, where the slow mode $\phi^\mu$ reaches its equilibrium value $\bar\phi^\mu$, the entropy density is maximized and equal to its standard equilibrium value
\begin{align}
\label{eq:hydro+_1st_law}
    &s(e,n)=\max_{\phi} s_{(+)}(e,n,\phi^\mu)=s_{(+)}(e,n,\bar\phi^\mu)\,,\nonumber\\
   &\pi^\mu(e,n,\bar\phi^\mu)=0\,.
\end{align}
In other words, deviations of the slow mode $\phi^\mu$ from its equilibrium value $\bar\phi^\mu$ reduce the entropy for a given hydrodynamic cell.

We also need to supply the evolution equation for $\phi^\mu$, which describes its relaxation to the equilibrium value $\bar\phi^\mu$ (cf. Ref.~\cite{Stephanov:2017ghc} for a scalar field). Using the same notation as in Eq.~(\ref{eq:IS_nmu}) we write
\begin{equation}
\label{eq:hydro+eq}
    \dot{\phi}^{\langle\mu\rangle} = - F^\mu_\phi + A^\mu_\nu\nabla^\nu\alpha_{(+)} + \cdots\,.
\end{equation}
The external force $\nabla\alpha_{(+)}$ drives the slow mode away from equilibrium, $F^\mu_\phi$ is a ``returning force'' driving $\phi^\mu$ back to equilibrium (thus $F^\mu_\phi{\,=\,}0$ when $\phi^\mu{\,=\,}\bar\phi^\mu$ and hence, according to \eqref{eq:hydro+_1st_law}, $\pi^\mu{\,=\,}0$), and $A^\mu_\nu$ is the (inverse) susceptibility of $\phi^\mu$ to the external force $\nabla\alpha_{(+)}$. Eq.~\eqref{eq:hydro+eq} requires $u_\mu A^\mu_\nu=0$, thus we can introduce a scalar $A_\phi$ so that $A^\mu_\nu=A_\phi\Delta^\mu_\nu$. The form of $F^\mu_\phi$ and $A_\phi$ shall be specified below. In general, $\phi^\mu$ can also change in response to additional external forces (indicated by $\cdots$ in Eq.~\eqref{eq:hydro+eq}), such as collective expansion of the background (the $\theta$-term in Eq.~\eqref{eq:IS_nmu3} arising from ${\cal J}^\mu$), long-range electromagnetic fields, etc. Here, we focus on its change in response to the gradient of chemical potential in units of the temperature, $\alpha_{(+)}$, as relevant for our study of baryon diffusion.

The generalized (partial-equilibrium) entropy current is given by
\begin{equation}\label{eq:s_(+)}
    s^\mu_{(+)}=s_{(+)}u^\mu+\Delta s^\mu\,,
\end{equation}
where $\Delta s^\mu$ describes a spatial non-equilibrium entropy {\it current} in the local rest frame. Using hydrodynamic equations we arrive at
\begin{eqnarray}
\label{eq:div_s_(+)}
    d\cdot s_{(+)}&=&\bigl(s_{(+)}-\beta_{(+)}w+\alpha_{(+)}n\bigr)\theta+\beta_{(+)}\partial_\mu u_\nu\Pi^{\mu\nu}\nonumber\\
    &&-\pi\cdot F_\phi+\left(A_\phi\pi_\nu-n_\nu\right)\nabla^\nu\alpha_{(+)}\nonumber\\
    &&+d_\mu\left(\Delta s^\mu+\alpha_{(+)}n^\mu\right)\,,
\end{eqnarray}
where Eqs.~\eqref{eq:gener_1st} and \eqref{eq:hydro+eq} were used.
The second law of thermodynamics requires the above expressions to be positive semidefinite, resulting in the following constraints: 
\begin{subequations}
\begin{eqnarray}
    s_{(+)}&=&\beta_{(+)}w-\alpha_{(+)}n\,,\\
    F^\mu_\phi&=&\gamma_\pi\pi^\mu,\\
    n^\mu&=&\kappa_{n(+)}\nabla^\mu\alpha_{(+)}+A_\phi\pi^\mu,\label{eq:hydro+_constraint_diffusion}\\
    \Delta s^\mu&=&-\alpha_{(+)}n^\mu.\label{eq:Delta_s_mu}
\end{eqnarray}
\end{subequations}
Here $\gamma_\pi$ and $\kappa_{n(+)}$ are (positive semidefinite) transport coefficients. \ld{We note that in Eq.~\eqref{eq:hydro+_constraint_diffusion} $A_\phi\pi^\mu$ corresponds to the contribution to $n^\mu$ arising from $\phi^\mu$. \xa{Thus $\kappa_{n(+)}$ amounts to the baryon transport coefficient in the absence of $\phi^\mu$ (i.e., $\kappa_{n(+)}=\kappa_{n,0}$), and $n^\mu$ approaches the conventional Navier-Stokes limit when $\phi^\mu$ is ignored.}} The constraint on $\Pi^{\mu\nu}$ (not displayed) leads to its conventional Navier-Stokes form. The last term in (\ref{eq:div_s_(+)}) can in general have either sign; to always satisfy the second law of thermodynamics we must require it to vanish, giving rise to Eq.~\eqref{eq:Delta_s_mu}.

The extended entropy can be decomposed as
\begin{equation}
    s_{(+)}(e, n, \phi^\mu)=s(e,n)+\Delta s(e, n, \phi^\mu)\,,
\end{equation}
where we postulate the longitudinal entropy correction due to the mode $\phi^\mu$ as (motivated by, e.g., Ref.~\cite{Israel:1979}) 
\begin{equation}
\label{eq:Delta_s}
    \Delta s=\frac{1}{2}\pi_\mu\phi^\mu\equiv\frac{1}{2}\pi_\phi\phi_\mu\phi^\mu,
\end{equation}
where $\pi_\phi$ is the susceptibility. Eq.~\eqref{eq:Delta_s} is quadratic in the space-like vector $\phi^\mu$ and always decreases the entropy. \ld{Using the Landau-Khalatnikov formula, the susceptibility $\pi_\phi$ can be rewritten as}
\begin{equation}
\label{eq:A_phi}
    1/\pi_\phi=A_\phi=\kappa_\ld{\phi}\Gamma_\phi\,,
\end{equation}
where $\Gamma_\phi=\gamma_\pi\pi_\phi$ is the relaxation rate \ld{of $\phi^\mu$, and $\kappa_\phi$ is introduced as a new transport coefficient whose physical meaning shall become clear right away}. Substituting the above expressions into Eq.~\eqref{eq:hydro+eq} \ld{and ignoring \xa{other} external force terms} we find
\begin{equation}
\label{eq:Dphi}
    \ld{\dot{\phi}^{\langle\mu\rangle} = - \Gamma_\phi \left(\phi^\mu-\kappa_\phi\nabla^\mu\alpha_{(+)}\right).}
\end{equation}
\ld{Now one can choose $\phi^\mu$ to represent the slowest mode contributing to $\n^\mu$, i.e., $G_{m\mu}\sim\langle\delta m \delta u_\mu \rangle$ and, near the critical point, think of $\phi^\mu$ as the ``critical sector'' of the baryon diffusion current. As discussed in Sec.~\ref{sec:critical_scaling}, for $\phi^\mu\sim G_{m\mu}$ with a typical wave number $q\sim\xi^{-1}$, $\Gamma_\phi\sim\xi^{-2}$ and $\kappa_\phi\sim\xi$; near the critical point, it controls the relaxation \xa{of the slowest mode contributing to the baryon diffusion}. 

Considering the Navier-Stokes limit of the baryon diffusion given by Eq.~\eqref{eq:hydro+_constraint_diffusion}, we can write down the relaxation equation for $n^\mu$. It receives a contribution of typical wave number $q\sim\xi^{-1}$ from $\phi^\mu$,
\begin{equation}\label{eq:NS_nmu}
    \dot{n}^{\langle\mu\rangle} = - \Gamma_n \left(n^\mu-\kappa_{n}\nabla^\mu\alpha_{(+)}\right),
\end{equation}
where we have used the fact that $\Gamma_n\simeq\Gamma_\phi\sim\xi^{-2}$ near the critical point. Here $\kappa_{n}\equiv\kappa_{n,0}+\kappa_\phi$, with $\kappa_{n,0}$ denoting the non-critical value of the baryon transport coefficient. Near the critical point $\kappa_\phi\sim\xi$ dominates $\kappa_n$ and consequently $\kappa_n\simeq\kappa_\phi\sim\xi$. The parametrization in Eq.~\eqref{eq:chi_kappa_scaling} is designed to reproduce this behavior approximately. Finally $\alpha_{(+)}$ is the normalized chemical potential with modifications from $\phi^\mu$ which generate critical behavior in the Equation of State near the critical point.
}
One sees that the single-mode Hydro+ equation \eqref{eq:NS_nmu} (\ld{which includes only a single wave number $q\sim\xi^{-1}$}) matches the Israel-Stewart type equation \eqref{eq:IS_nmu2} for $n^\mu$ \ld{when the critical scaling (\ref{eq:chi_kappa_scaling},\ref{eq:tau_n_xi}) in the critical regime is accounted for.}\footnote{\uh{The last term in Eq.~\eqref{eq:IS_nmu2}, which is $\sim \theta n^\mu$, is subdominant in the critical regime.}}

Eq.~(\ref{eq:NS_nmu}) can be solved in frequency ($\omega$) space as
\begin{equation}\label{eq:phi^mu_omega}
    n^\mu(\omega)=\kappa_n(\omega)\nabla^\mu\alpha_{(+)}\,,
\end{equation}
where
\begin{equation}\label{eq:kappa_omega}
    \kappa_n(\omega)\equiv f_\kappa\bigl(\omega/\Gamma_n\bigr)\,\kappa_n
\end{equation}
is the frequency-dependent baryon diffusion coefficient, with
\begin{equation}\label{eq:f_kappa}
    f_\kappa(x)=\frac{1+ix}{1+x^2}\,,
\end{equation}
and $\kappa_n{\,\equiv\,}\kappa_n(\omega{=}0)$ being the frequency-independent part. While the imaginary part of Eq.~\eqref{eq:f_kappa} relates the baryonic analog of the electric permittivity, its real part gives rise to the frequency dependence of the baryon transport coefficient:
\begin{equation}
\label{eq:kappa_frequency}
    \text{Re} [\kappa_n(\omega)]=\text{Re} [f_\kappa\bigl(\omega/\Gamma_n\bigr)]\,\kappa_n=\frac{\kappa_n}{1+(\omega/\Gamma_n)^2}\,.
\end{equation}
One can infer from Eq.~\eqref{eq:kappa_frequency} that at small $\omega$, \xa{$\kappa_n(\omega)-\kappa_n(0)\sim\ld{\kappa_n(0)}\,\omega^2$} while at large $\omega$, $\kappa_n(\omega)\sim\ld{\kappa_n(0)}\,\omega^{-2}$, different from the results given in Ref.~\cite{An:2019hydro++}.\footnote{\label{fn6}%
    In an analysis that takes modes with all wave numbers into account \cite{An:2019hydro++}, \xa{$\kappa_n(\omega)-\kappa_n(0) \sim \ld{\kappa_n(0)}\omega^{1/2}$} at small $\omega$, and $\kappa_n(\omega) \sim \ld{\kappa_n(0)}\omega^{-1/2}$ at large $\omega$. The non-integer power indicates the non-analytic behavior of the long-time tails.}

Eqs.~\eqref{eq:phi^mu_omega}-\eqref{eq:kappa_omega} generalize the Navier-Stokes solution for the baryon diffusion current into the critical region where, due to critical slowing down, the slow critical fluctuation modes are not in thermal equilibrium. That is to say, these equations indicate that at large frequencies $\omega/\Gamma_n{\,\gtrsim\,}1$ baryon diffusion is suppressed, hence a naive extrapolation of hydrodynamics with the frequency-independent coefficient $\kappa_n(0){\,\sim\,}\xi$ (see Eq.~\eqref{eq:coeff_scaling}) to this regime would overestimate the amount of baryon diffusion. Knowing from Eq.~\eqref{eq:tau_n_xi} that $\Gamma_n{\,\sim\,}\xi^{-2}$, this further implies that the suppression affects modes with frequencies $\xi^{-2} \lesssim \omega \ll \xi^{-1}$, i.e. inside the Hydro++ regime discussed in Ref.~\cite{An:2019hydro++}. When analyzed within the Hydro+ framework with only a single slow mode, the switching-off \ld{of critical contributions to baryon diffusion} occurs at $\omega\,\xi^2\sim 1$; this is consistent with the general analysis in Ref.~\cite{An:2019hydro++} where the full \xa{wave number} spectrum is taken into account. In other words, \xa{the switching-off of \ld{the critical contribution to} baryon diffusion} at finite frequency is taken care of by Eq.~\eqref{eq:NS_nmu} (and similarly by Eq.~\eqref{eq:IS_nmu2}). 

\ld{Summarizing briefly,} we emphasize that in the critical regime the Hydro+ equation \eqref{eq:NS_nmu} leads to simi\-lar dynamics as the Israel-Stewart equation \eqref{eq:IS_nmu2} for the diffusion current $n^\mu$ since both equations correctly account for critical slowing \ld{down} through $\Gamma_n=1/\tau_n$. This correspondence is expected since in the critical regime the off-equilibrium effects are dominated by fluctuations that are effectively frozen. However, Eq.~\eqref{eq:NS_nmu} still differs from the Hydro++ equations presented in Ref.~\cite{An:2019hydro++}, since only a single representative mode (with the typical wave number $q\sim\xi^{-1}$) is analyzed. The \xa{exact asymptotic suppression behavior} as well as the non-analytic frequency dependence of $\kappa_n$ (cf. footnote~\ref{fn6}) arising in Hydro++ from the phase-space integration over all critical modes are both missed by Eq.~\eqref{eq:NS_nmu}. Indeed, the critical effects on $\kappa_n$ are overestimated by Eq.~\eqref{eq:NS_nmu} at small $\omega$ (i.e., $\kappa_n$ is less suppressed compared to Ref.~\cite{An:2019hydro++}). However, we shall see in the following that the resulting overestimation of critical effects is negligible when compared to much stronger suppression effects arising from different origins. For these reasons the single-mode Hydro+ \ld{formalism (or, equivalently, the generalized Israel-Stewart formalism amended by critical scaling)} serves as a good prototype of the state-of-the-art Hydro+/++ theory -- it is sophisticated enough to capture the phenomenologically important feature of critical slowing down while preserving computational economy.

\section{Setup of the framework}\label{sec:setup}

In this section we set up the framework for simulating the evolution of a fireball close to the QCD critical point. The core of our framework is the hydrodynamic equations discussed in Sec.~\ref{sec:hydro}. This requires specification of the EoS and the transport coefficients, as well as  initial and final conditions. We also discuss the  particlization process of converting the fluid dynamic output into particles whose momentum distributions (after integrating over the conversion hypersurface) can be compared with experimental measurements.

\textbf{\textit{Initial conditions.}} We start with the initial conditions which, from a physics perspective, describe the initial state of the systems while mathematically providing the initial data for solving the initial value problem associated with our coupled set of partial differential equations. At collision energies of order $\snn\sim{}O(10)\,$GeV, the longitudinal interpenetration dynamics of the two colliding nuclei becomes complicated, in principle requiring a time-dependent, (3+1)-dimensional description of the initial energy-momentum deposition and baryon number doping processes that produce the QGP fluid. Recently, several so-called ``dynamical initialization'' algorithms have been proposed to address this problem (see Refs.~\cite{Shen:2017ruz, Shen:2017bsr, Akamatsu:2018olk, Du:2018mpf} and the reviews \cite{Shen:2020gef, Shen:2020mgh}). In this exploratory study, however, we try to establish a basic understanding of baryon diffusion dynamics for Au-Au collisions at $\snn\eq19.6$\,GeV in which we focus entirely on the longitudinal dynamics, modeling a (1+1)-dimensional system without transverse gradients initiated instantaneously at a constant proper time $\tauI$ (see also Refs.~\cite{Monnai:2016kud, Li:2018fow}). More specifically, we evolve the system hydrodynamically using the longitudinal initial profiles $e(\tauI, \eta_s),\, n(\tauI, \eta_s)$ provided in Ref.~\cite{Denicol:2018wdp}, starting at $\tauI\eq1.5$\,fm/$c$.\footnote{%
    Ref.~\cite{Denicol:2018wdp} provides longitudinal initial distributions for the entropy and baryon densities. We here adopt the functional form of their initial entropy profile as our energy profile, after appropriate normalization.}
The initial hydrodynamic profiles are shown as gray curves in Fig.~\ref{fig:long_evolution} below. The initial energy density has a plateau covering the space-time rapidity $\eta_s\in[-3.0,\,3.0]$ whereas the initial net baryon density features a double peak structure and covers a narrower region  $\eta_s\in[-2.0,\,2.0]$, reflecting baryon stopping.\footnote{%
    We note that baryon stopping affects the initial momentum rapidity $y$ of the baryon number carrying degrees of freedom and is typically modelled by a rapidity shift $\Delta y\sim 1-1.5$, depending on system size and collision energy. To translate this rapidity shift into a shift in space-time rapidity $\eta_s$ (as done in Fig.~\ref{fig:long_evolution}) requires a dynamical initialization model. Different such models yield different initial density and flow profiles \cite{Shen:2017ruz, Shen:2017bsr, Akamatsu:2018olk, Du:2018mpf, Shen:2020gef, Shen:2020mgh}.}
For the initial longitudinal momentum flow we take the ``static'' flow profile $u^\mu = (1,0,0,0)$ in Milne coordinates (corresponding to Bjorken expansion \cite{Bjorken:1982qr} in Cartesian coordinates), and the initial baryon diffusion current is assumed to vanish, $n^\mu = (0,0,0,0)$. 

\textbf{\textit{EoS and transport coefficients.}} Given these initial conditions, the hydrodynamic equations \eqref{eq:conservation} and \eqref{eq:IS_nmu3} are solved by \beshydro\ \cite{Du:2019obx}.
For the EoS at non-zero net baryon density we use {\sc neos} \cite{Monnai:2019hkn} which was constructed by smoothly joining Lattice QCD data \cite{Bazavov:2014pvz, Bazavov:2012jq, Ding:2015fca, Bazavov:2017dus} with the hadron resonance gas model. As already mentioned, we here focus on baryon diffusion dynamics by ignoring shear and bulk viscous stresses. For the transport coefficients related to baryon diffusion we rely on the theoretical work in Refs.~\cite{Son:2006em, Rougemont:2015ona, Denicol:2018wdp, Soloveva:2019xph} since phenomenological constraints are still lacking. Specifically, we here use the coefficients obtained from the Boltzmann equation for an almost massless classical gas in the relaxation time approximation (RTA) \cite{Denicol:2018wdp}, which gives for the baryon diffusion coefficient
\begin{equation}
\label{eq:kappa_kinetic}
    \kappa_{n,0} = C_n\frac{n}{T}\,\left(\frac{1}{3}\coth\alpha-\frac{n T}{w} \right)\,,
\end{equation}
and for the relaxation time
\begin{equation}
\label{eq:taun_kinetic}
    \tau_{n,0}= \frac{C_n}{T}\,,
\end{equation}
where $C_n$ is a free unitless parameter.\footnote{%
    In Israel-Stewart theory \cite{Israel:1979}  $\tau_{n,0}=\lambda_{T,0} T \beta_n$ where $\beta_n$ is a second-order transport coefficient and $\lambda_{T,0}$ is the non-critical thermal conductivity associated with $\kappa_{n,0}$ through Eq.~\eqref{eq:kappa-lambda}.} 
Throughout this paper, we set $C_n\eq0.4$; in  Ref.~\cite{Denicol:2018wdp} this value was shown to yield good agreement with selected experimental data. Following the kinetic theory approach \cite{Denicol:2018wdp} we also set $\delta_{nn,0}=\tau_{n,0}$ in Eq.~(\ref{eq:IS_nmu3}) as its non-critical value. In the limit of zero net baryon density, $\kappa_{n,0}$ remains non-zero at non-zero temperature, $\lim_{\mu\to0} \kappa_{n,0}/\tau_{n,0} = nT/(3\mu)$ \cite{Denicol:2018wdp} --- a feature also seen in  holographic models; for example, using the AdS/CFT correspondence, the (baryon) charge conductivity of $r$-charged black holes translates into \cite{Son:2006em,Li:2018fow}
\begin{equation}\label{eq:kappa_holo}
    \kappa_{n,0} = 2\pi\frac{Ts}{\mu^2}\left(\frac{nT}{w}\right)^2\,.
\end{equation}%
%
\begin{figure}[!tbp]
\begin{center}
\hspace{-1cm}
\includegraphics[width=0.38\textwidth]{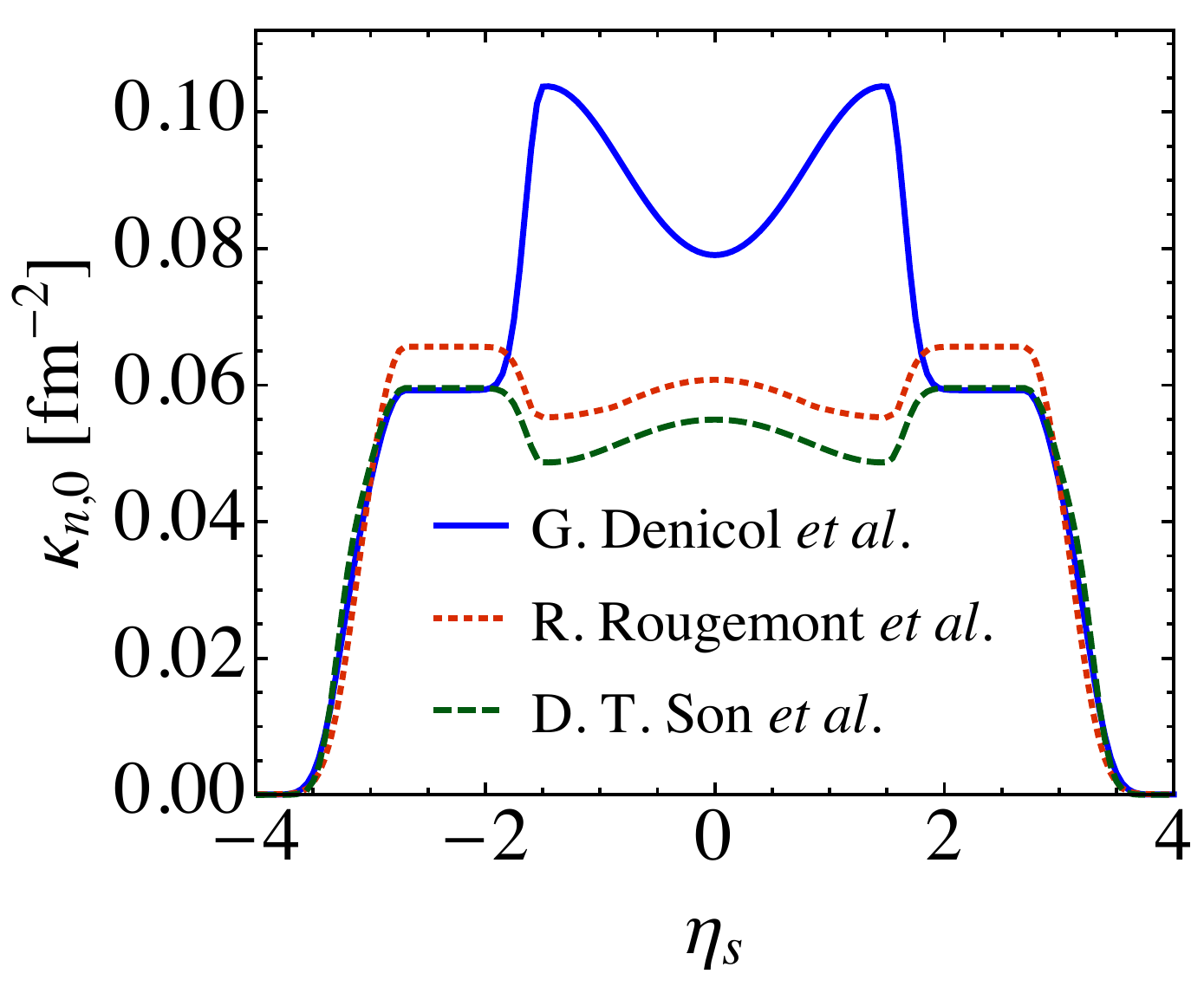}
\caption{%
    Initial longitudinal distributions of $\kappa_{n,0}$, corresponding to the initial profiles in Fig.~\ref{fig:long_evolution} below. $\kappa_{n,0}$ is calculated using different methods, including kinetic theory \cite{Denicol:2018wdp} (Eq.~\eqref{eq:kappa_kinetic}, blue solid line) and two holographic models, using Eq.~\eqref{eq:kappa_holo} (Ref.~\cite{Son:2006em}, green dashed line) and tabulated values found in Refs.~\cite{Rougemont:2015ona, Rougemont:2015wca} (red dotted line).
    \label{fig:initial_kappa_tau}}
\end{center}
\end{figure}%
%

Since $\kappa_{n,0}$ is such an important parameter in our study, we offer some intuition about its key characteristics in Fig.~\ref{fig:initial_kappa_tau}, where its initial space-time rapidity profile is plotted for three of the theoretical approaches referenced above.\footnote{%
    Ref.~\cite{Li:2018fow} compared the expressions of $\kappa_{n,0}$ in Eqs.~(\ref{eq:kappa_kinetic}) and (\ref{eq:kappa_holo}) while a comparison of Eq.~(\ref{eq:kappa_kinetic}) with a different holographic approach \cite{Rougemont:2015ona, Rougemont:2015wca} was previously presented in Ref.~\cite{Du:2018mpf}.
} 
\ld{The figure shows that in the region $|\eta_s|\lesssim2$, where the net baryon density is non-zero (cf.~Fig.~\ref{fig:long_evolution}d below), differences exist in $\kappa_{n,0}(\eta_s)$ between the weakly and strongly coupled approaches: In the holographic approaches $\kappa_{n,0}$ is {\em suppressed} by baryon density while in the kinetic approach it is {\em enhanced}. On the other hand we see that at large rapidity $|\eta_s|\gtrsim2.5$, where the baryon density approaches zero (cf.~Fig.~\ref{fig:long_evolution}d), the different models for $\kappa_{n,0}$ yield similar distributions, all of them rapidly decreasing towards zero as the temperature decreases (cf.~Fig.~\ref{fig:long_evolution}b). This also implies that generically, as the fireball expands and cools down, all three approaches yield rapidly decreasing amounts of baryon diffusion, as will be discussed below in Sec.~\ref{sec:tevol}.}

\begin{figure*}[!tbp]
\begin{center}
\hspace{-1cm}
\includegraphics[width=1.05\textwidth]{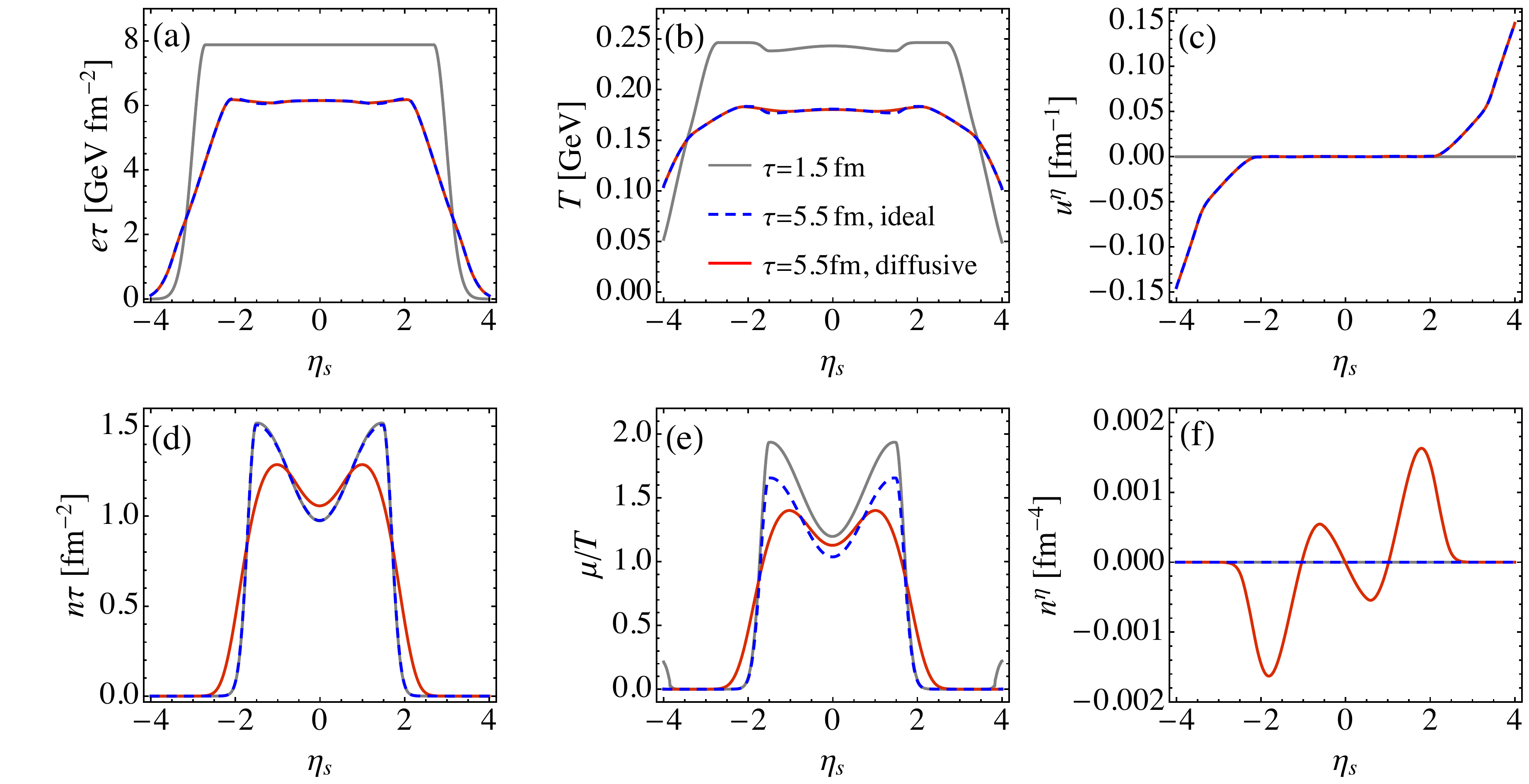}
\caption{%
    Longitudinal evolution for the cases without (``ideal'', blue dashed line) and with baryon diffusion (``diffusive'', red solid line). The gray curves show the initial distributions at the initial time $\tauI\eq1.5$\,fm/$c$; colored curves show the evolved distributions at $\tau\eq5.5$\,fm/$c$ for ideal and diffusive evolution, respectively. Note that both energy and baryon densities are scaled by the proper time $\tau$.}
    \label{fig:long_evolution}
\end{center}
\end{figure*}

\textbf{\textit{Particlization.}} After completion of the hydrodynamic evolution (results of which will be discussed in Sec.~\ref{sec:results}) we compute the particle distributions corresponding to the hydrodynamic fields on the freeze-out surface $\Sigma$, using the Cooper-Frye formula \cite{Cooper:1974mv}:
\begin{eqnarray}
    E \frac{d^3 N_i}{d^3 p} = \frac{g_i}{(2\pi)^3}\int_\Sigma p^\mu d^3 \Sigma_\mu (f_{\mathrm{eq},i} + \delta f_{\mathrm{diff},i})\,,
\label{eq:CooperFrye}
\end{eqnarray}
where $E,\bm{p}$ are the energy and momentum of a particle of species $i$ in the observer frame, $g_i$ is the spin-isospin degeneracy factor, $d^3 \Sigma_\mu(x)$ is the outward-pointing surface normal vector at point $x$ on the three-surface $\Sigma$, $f_{\mathrm{eq},i}$ is the equilibrium distribution for particle of species $i$, and $\delta f_{\mathrm{diff},i}$ the off-equilibrium correction resulting from net baryon diffusion (other viscous corrections are neglected in this paper). At first order in the Chapman-Enskog expansion of the RTA Boltzmann equation, the dissipative correction from net baryon diffusion $\delta f_{\mathrm{diff},i}$ is given by \cite{Denicol:2018wdp, McNelis:2019auj, McNelis:2021acu}
\begin{eqnarray}
    \delta f_{\mathrm{diff},i} = f_{\mathrm{eq},i} (1 -\theta_i f_{\mathrm{eq},i}) \left(\frac{n}{w} - \frac{b_i}{E} \right) \frac{p^{\langle \mu \rangle} n_\mu}{\hat{\kappa}}\,,
\label{eq:diffusion_deltaf}
\end{eqnarray}
where $\theta_i\eq1$ $(-1)$ for fermions (bosons), $b_i$ is the baryon number of particle species $i$, $p^{\langle \mu \rangle}\equiv\Delta^{\mu\nu}p_\nu$, and $\hat{\kappa}=\kappa_n/\tau_n$. We will discuss how the critical correction is included, as well as its effects on the final particle distributions, in Sec.~\ref{sec:diffcp_spectra}. We evaluate the continuous momentum distribution (\ref{eq:CooperFrye}) numerically using the {\sc iS3D} particlization module \cite{McNelis:2019auj}, ignoring rescattering among the particles and resonance decays after particlization.

\section{Results and discussion}
\label{sec:results}

In this section we discuss the dynamics of the fireball at a fixed collision energy of $\snn\eq19.6$\,GeV. First we study in Sec.~\ref{sec:diffnocp} baryon diffusion effects on its $T$-$\mu$ trajectories through the phase diagram for cells located at different space-time rapidities \etas, and in Sec.~\ref{sec:diffnocp_particles} on freeze-out surface and final particle distributions, in the absence of critical dynamics. Then, in Sec.~\ref{sec:diffcp}, we discuss how the critical behavior described in Sec.~\ref{sec:critical} modifies this dynamics for cells whose trajectories pass close to the critical point. In Sec.~\ref{sec:tevol} we point out some generic features of the time evolution of baryon diffusion.

\subsection{Longitudinal dynamics of baryon evolution}
\label{sec:diffnocp}

In Figure~\ref{fig:long_evolution}, we show snapshots of the longitudinal distributions of the hydrodynamic quantities at two times, the initial time 1.5\,fm/$c$ (gray solid lines) and later at $\tau\eq5.5$\,fm/$c$, with and without baryon diffusion (red solid and blue dashed lines, respectively). The gray curves in panels (a) and (d)  show the initial energy and baryon density distributions from Ref.~\cite{Denicol:2018wdp}. The gray lines in panels (b) and (e) show the corresponding temperature and chemical potential profiles, extracted with the {\sc neos} equation of state \cite{Monnai:2019hkn, Bazavov:2014pvz, Bazavov:2012jq, Ding:2015fca, Bazavov:2017dus}. The gray horizontal lines in panels (c) and (f) show the zero initial conditions for the longitudinal flow and baryon diffusion current. The temperature profile (panel (b)) shares the plateau with energy density (panel (a)), up to small structures caused by the double-peak structure of the baryon density (panel (d)) and baryon chemical potential profiles (panel (e)). The chemical potential in panel (e) inherits the double-peak structure from baryon density in panel (d). These structures are also reflected in the pressure (not shown).\footnote{%
    In the very dilute forward and backward rapidity regions 
    one observes a steep rise of the initial $\mu/T$. This feature is sensitive to the rates at which $e$ and $n$ approach zero as $|\eta_s|\to\infty$, and it is easily affected by numerical inaccuracies. Since both $T$ and $\mu$ are close to zero there, the baryon diffusion coefficient $\kappa_{n,0}$ also vanishes, and (as seen in Fig.~\ref{fig:long_evolution}f) the apparently large but numerically unstable gradient of $\mu/T$ at large $\eta_s$ does not generate a measurable baryon diffusion current.}

We next discuss the blue dashed lines in Fig.~\ref{fig:long_evolution} showing the results of ideal hydrodynamic evolution. Work done by the longitudinal pressure converts thermal energy into collective flow kinetic energy such that the thermal energy density $e$ decreases faster than $1/\tau$ (panel (a)). Small pressure variations along the plateau of the distribution (caused by the rapidity dependence of $\mu/T$) lead to slight distortions of the rapidity plateau of the energy density as its magnitude decreases. Longitudinal pressure gradients at the forward and backward edges of the initial rapidity plateau accelerate the fluid longitudinally, generating a non-zero $\eta_s$-component of the hydrodynamic flow at large rapidities (panel (c)). As seen in panels (a) and (c), the resulting longitudinal rarefaction wave travels inward slowly, leaving the initial Bjorken flow profile $u^\eta\eq0$ untouched for $|\eta_s|<2.5$ up to $\tau\eq5.5$\,fm/$c$. For Bjorken flow without transverse dynamics baryon number conservation implies that $n\tau$ remains constant. Panel (d) shows this to be the case up to $\tau\eq5.5$\,fm/$c$ because, up to that time, the initial Bjorken flow has not yet been affected by longitudinal acceleration over the entire $\eta_s$-interval in which the net baryon density $n$ is non-zero. Panel (e) shows, however, that in spite of $n\tau$ remaining constant within that $\eta_s$ range, the baryon chemical potential $\mu/T$ decreases with time, as required by the {\sc neos} equation of state.

The nontrivial evolution effects of turning on the baryon diffusion current via Eq.~(\ref{eq:IS_nmu3}) are shown by the red solid lines in Fig.~\ref{fig:long_evolution}. The baryon diffusion current itself is plotted in panel (f) and will be discussed shortly. Panels (a)-(c) show that baryon diffusion has almost no effect at all on the energy density  (and, by implication, on the pressure), the temperature, and the hydrodynamic flow generated by the pressure gradients. Given the weak dependence of pressure and temperature on baryon density through the EoS this is to be expected. Baryon diffusion does, however, significantly modify the rapidity profiles of the net baryon density (d), chemical potential $\mu/T$ (e). Generated by the negative gradient of $\mu/T$, the baryon diffusion current moves baryon number from high- to low-density regions, causing an overall broadening of the baryon density rapidity profile in (d) while simultaneously filling in the dip at midrapidity \cite{Shen:2017ruz, Denicol:2018wdp, Li:2018fow, Du:2018mpf, Du:2019obx, Fotakis:2019nbq}. Panel (e) shows how the chemical potential $\mu/T$ tracks these changes in the baryon density profile (panel (d)), and panel (f) shows the baryon diffusion current responsible for this transport of baryon density, with its alternating sign and magnitude tracing the sign and magnitude changes of $-\nabla(\mu/T)$. As we shall see in Sec.~\ref{sec:tevol}, the smoothing of the gradients of baryon density and chemical potential contributes to a fast decay of baryon diffusion effects.

\begin{figure}[!tbp]
\begin{center}
\hspace{-.5cm}
\includegraphics[width=0.4\textwidth]{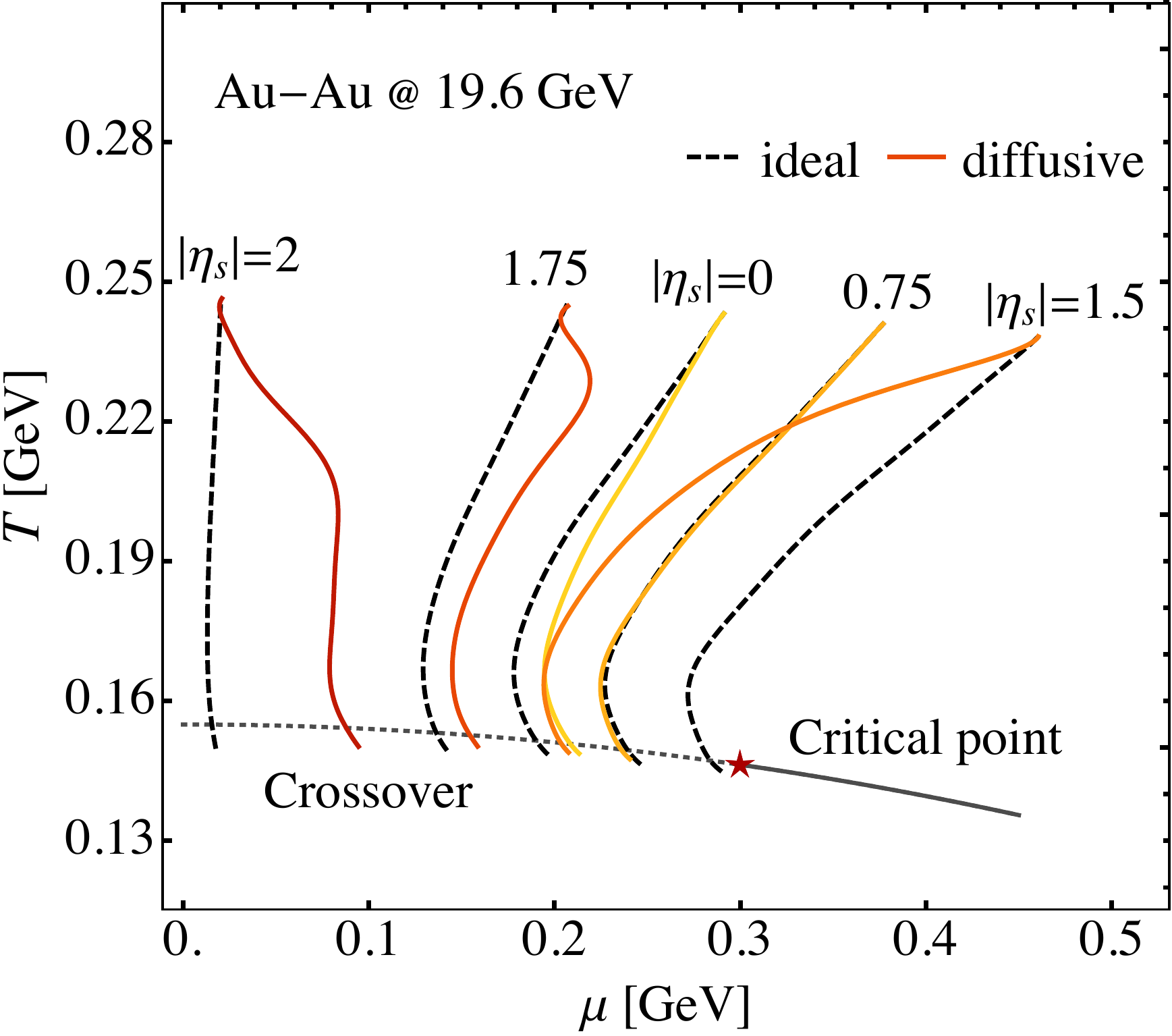}    
\caption{%
    Phase diagram trajectories of fluid cells at different $|\eta_s|$ for the Au+Au collision fireball discussed here.  Black dashed lines indicate ideal evolution while colored solid lines include the effects of baryon diffusion. All fluid cells evolve from high to low temperature. The phase transition line and critical point are included only to guide the eye -- critical effects on the EoS and transport coefficients are not included in this figure.}
    \label{fig:phase_dia_traj}
\end{center}
\end{figure}

Fig.~\ref{fig:long_evolution} indicates non-trivial thermal, chemical, and mechanical evolution at different rapidities. Fluid cells at different \etas{} pass through different regions of the QCD phase diagram and may therefore be affected differently by the QCD critical point \cite{Monnai:2016kud, Shen:2018pty, Dore:2020jye}. This has led to the suggestion \cite{Brewer:2018abr} of using rapidity-binned cumulants of the final net proton multiplicity distributions as the possibly sensitive observables of the critical point.\footnote{%
    We caution that at BES energies the mapping between space-time rapidity \etas{} of the fluid cells and rapidity $y$ of the emitted hadrons is highly nontrivial and requires dynamical modelling.}
To illustrate the point we show in Fig.~\ref{fig:phase_dia_traj} the phase diagram trajectories of fluid cells at several selected $|\eta_s|$ values,\footnote{%
    Cells at opposite but equal space-time rapidities are equivalent because of $\eta_s\to-\eta_s$ reflection symmetry in this work.}
both with and without baryon diffusion. As we move from mid-rapidity to $|\eta_s|\eq2.0$, the starting point of these trajectories first moves from $\mu\simeq0.28$\,GeV at $\eta_s\eq0$ to the larger value $\mu\simeq0.45$\,GeV at $\eta_s\eq1.5$, but then turns back to $\mu\simeq0.2$\,GeV at $\eta_s\eq1.75$, and finally to $\mu\simeq0$ at $\eta_s\eq2.0$, without much variation of the initial temperature $T_i\simeq0.25$\,GeV (see Figs.~\ref{fig:long_evolution}b,e). The difference between the dashed (ideal) and solid (diffusive) trajectories exhibits a remarkable dependence on \etas{}: Both the sign and the magnitude of the diffusion-induced shift in baryon chemical potential depend strongly on space-time rapidity. In most cases, we note that the solid (diffusive) trajectories move initially rapidly away from the corresponding ideal ones, but then quickly settle on a roughly parallel ideal trajectory. A glaring exception is the trajectory of the cell at $\eta_s\eq1.5$, which starts at the maximal initial baryon chemical potential and keeps moving away from its initial ideal $T$-$\mu$ trajectory for a long period, settling on a new ideal trajectory only shortly before it reaches the hadronization phase transition. The reason for this behavior can be found in Fig.~\ref{fig:long_evolution}e, which shows that at $\eta_s\eq1.5$ the gradient of $\mu/T$ remains large throughout the fireball evolution. But almost everywhere else baryon diffusion effects die out quickly.

\begin{figure}[!tbp]
\begin{center}
\hspace{-0.5cm}
\includegraphics[width= 0.36\textwidth]{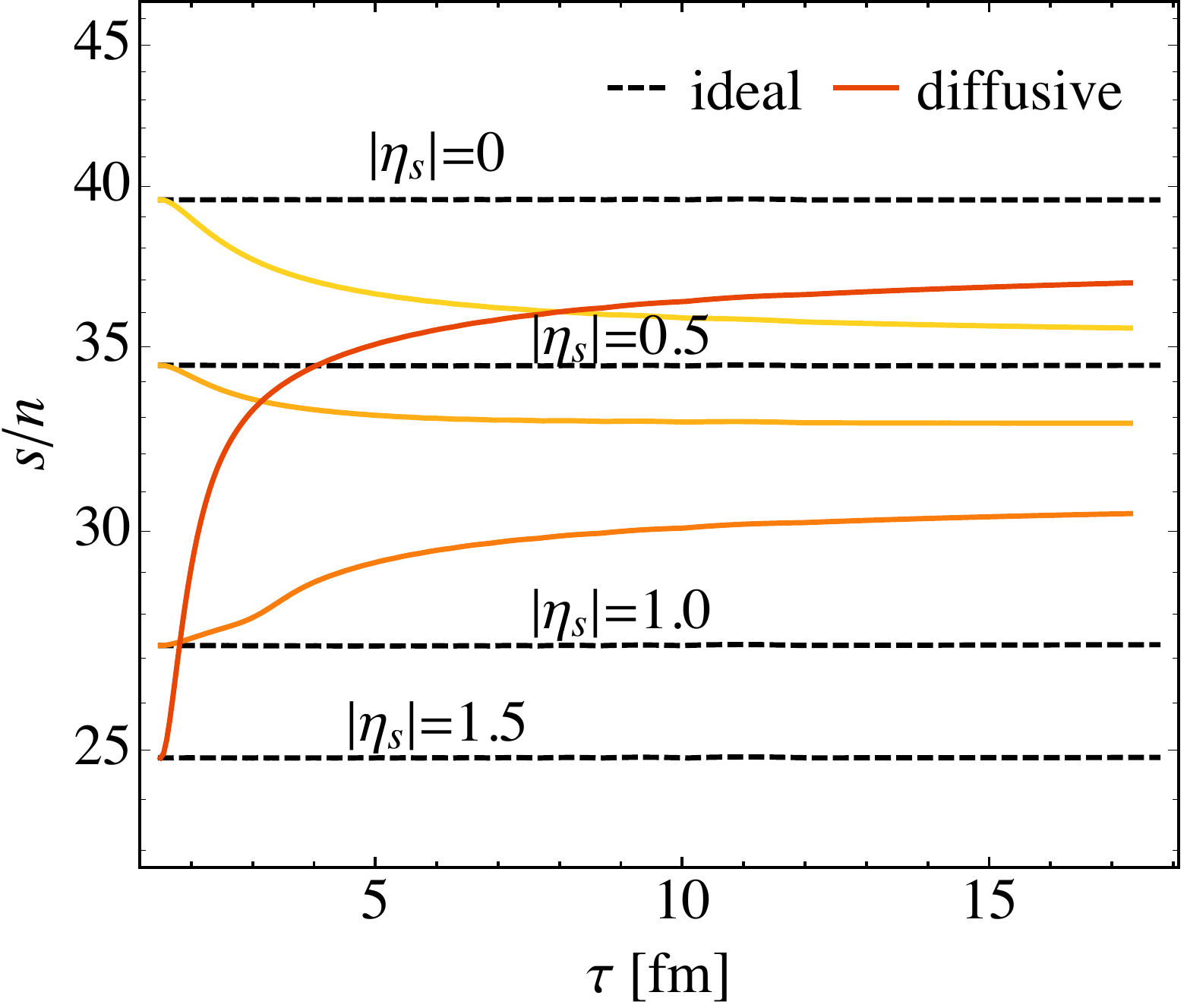}
\caption{%
    Time evolution of entropy per baryon at selected \etas{} values in ideal and diffusive fluid dynamics (dashed and solid lines, respectively).
    }
    \label{fig:sn_evolution}
\end{center}
\end{figure}

Since ideal fluid dynamics conserves both baryon number and entropy, the dashed trajectories are lines of constant entropy per baryon. This is shown by the dashed lines in Fig.~\ref{fig:sn_evolution}. Baryon diffusion leads to a net baryon current in the local momentum rest frame and thereby changes the baryon number per unit entropy. This is illustrated by the solid lines in Fig.~\ref{fig:sn_evolution}. Depending on the direction of the $\mu/T$ gradients, baryon diffusion can increase or decrease the entropy per baryon. 

We close this discussion by commenting on the turning of the dashed $m\equiv s/n\eq$const.{}~trajectories in Fig.~\ref{fig:phase_dia_traj} from initially pointing towards the lower left to later pointing towards the lower right. This is a well known feature of isentropic expansion trajectories in the QCD phase diagram \cite{Cho:1993mv, Hung:1997du, Monnai:2019hkn} that reflects the change in the underlying degrees of freedom, from quarks and gluons to a hadron resonance gas, at the point of hadronization as embedded in the construction of the EoS.

Figure~\ref{fig:phase_dia_traj} is reminiscent of the QCD phase diagram often shown to motivate the study of heavy-ion collisions at different collision energies in order to explore QCD matter at different baryon doping (see, for example, the 2015 DOE-NSF NSAC Long Range Plan for Nuclear Physics \cite{Geesaman:2015fha}). What had been shown there are (isentropic) expansion trajectories for matter created {\em at midrapidity in heavy-ion collisions with different beam energies}, whereas Fig.~\ref{fig:phase_dia_traj} shows similar expansion trajectories {\em for different parts of the fireball in a collision with a fixed beam energy}. Fig.~\ref{fig:phase_dia_traj} thus makes the point that in general the matter created in heavy ion collisions can never be characterized a single fixed value of $\mu/T$. At high collision energies space-time and momentum rapidities are tightly correlated, $\eta_s\simeq y$, and different $\eta_s$ regions with different baryon doping $\mu/T$ can thus be more or less separated in experiment by binning the data in momentum rapidity $y$. This motivates the strategy of scanning the changing baryonic composition in the $T$-$\mu$ diagram by performing a rapidity scan at fixed collision energy rather than a beam energy scan at fixed rapidity \cite{Monnai:2016kud, Shen:2018pty, Brewer:2018abr}. This strategy fails, however, at lower collision energies where particles of fixed momentum rapidity can be emitted from essentially every part of the fireball and thus receive contributions from regions with wildly different chemical compositions, with non-monotonic rapidity dependences that are non-trivially and non-monotonically affected by baryon diffusion.   

\subsection{Freeze-out surface and final particle distributions}
\label{sec:diffnocp_particles}

The expansion trajectories shown in the previous subsection all end at the same constant proper time (see Fig.~\ref{fig:sn_evolution}). In phenomenological applications it is usually assumed that the hydrodynamic stage ends and the fluid falls apart into particles when all fluid cells reach a certain ``freeze-out energy density'', here taken as $e_\mathrm{f}\eq0.3$\,GeV/fm$^3$.\footnote{%
    This is lower than the value of 0.4\,GeV/fm$^3$ used in Ref.~\cite{Denicol:2018wdp}, in order to ensure that the expansion trajectories reach into the hadronic phase below the crossover line from Ref.~\cite{Bellwied:2015rza}.}
With such a freeze-out criterion, fluid cells at different \etas{} freeze out at different times $\tau_\mathrm{f}(\eta_s)$. In this subsection we discuss this freeze-out surface and the distributions of particles emitted from it. 

\begin{figure}[!tbp]
\begin{center}
\hspace{-0.5cm}
\includegraphics[width= 0.48\textwidth]{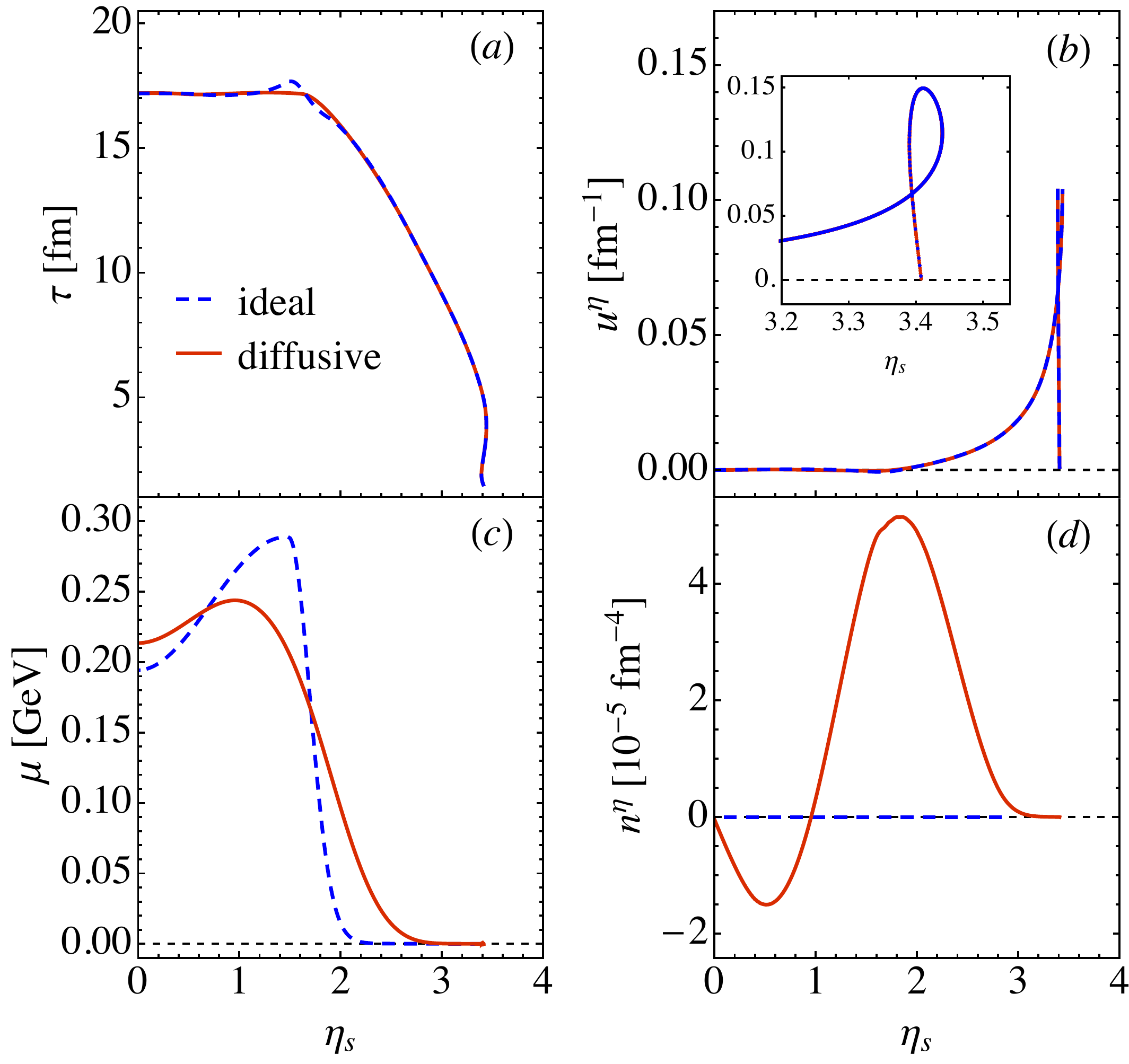}
\caption{%
    Distributions on the freeze-out surface, defined by a constant freeze-out energy density $e_\mathrm{f}\eq0.3\,$GeV/fm$^3$. Shown are as functions of \etas{} (a) the space-time profile of the freeze-out surface, (b) the longitinal flow, (c) the baryon chemical potential, and (d) the longitudinal component of the diffusion current. Blue dashed and solid red lines correspond to ideal and diffusive fluid dynamics, respectively. Note that in Milne coordinates the longitudinal flow velocity $u^\eta$ is not unitless --- $u^\eta \sqrt{-g} = \tau u^\eta$ is. Similar comments apply to the units of the baryon diffusion current $n^\eta$.}
    \label{fig:fz_surf}
\end{center}
\end{figure}

Fig.~\ref{fig:fz_surf} shows the freeze-out surface $\tau_\mathrm{f}(\eta_s)$ in panel (a) as well as the longitudinal flow, baryon chemical potential, and longitudinal component of the baryon diffusion current in panels (b)-(d).\footnote{%
    The freeze-out finder implemented in \beshydro{} is based on {\sc Cornelius} \cite{Huovinen:2012is}. We previously tested its efficiency within \beshydro{} at non-zero baryon density in the transverse plane \cite{du2021jet}. In App.~\ref{sec:long_evo_validation} we also validate it for this work, which features longitudinal dynamics with baryon diffusion current.}
Ideal and diffusive hydrodynamics are distinguished by blue dashed and red solid lines. Panel (a) shows that initially the longitudinal pressure gradient causes the fluid to grow in \etas{} direction before it starts to shrink after $\tau\gtrsim4$\,fm/$c$ due to cooling and surface evaporation. As seen in Fig.~\ref{fig:long_evolution}a, the core of the fireball remains approximately boost invariant while cooling by performing longitudinal work, until the longitudinal rarefaction wave reaches it. Once the energy density in this boost-invariant core drops below $e_\mathrm{f}$, it freezes out simultaneously, as seen in the flat top of the freezeout surface shown in panel (a). Slight deviations from boost invariance are caused by the effects of the boost-non-invariant net baryon density profile and its (small) effect on the pressure whose gradient drives the hydrodynamic expansion. Baryon diffusion has practically no effect on the freeze-out surface, nor on the longitudinal flow along this surface shown in panel (b), owing to the weak dependence of the EoS on baryon doping. The distributions of the baryon chemical potential and baryon diffusion current across this surface, on the other hand, are significantly affected by baryon diffusion, as seen in panels (c) and (d). It bears pointing out, however, that the magnitude of the baryon diffusion current in panel (d) is very small.

\begin{figure}[!tbp]
\begin{center}
\includegraphics[width= 0.36\textwidth]{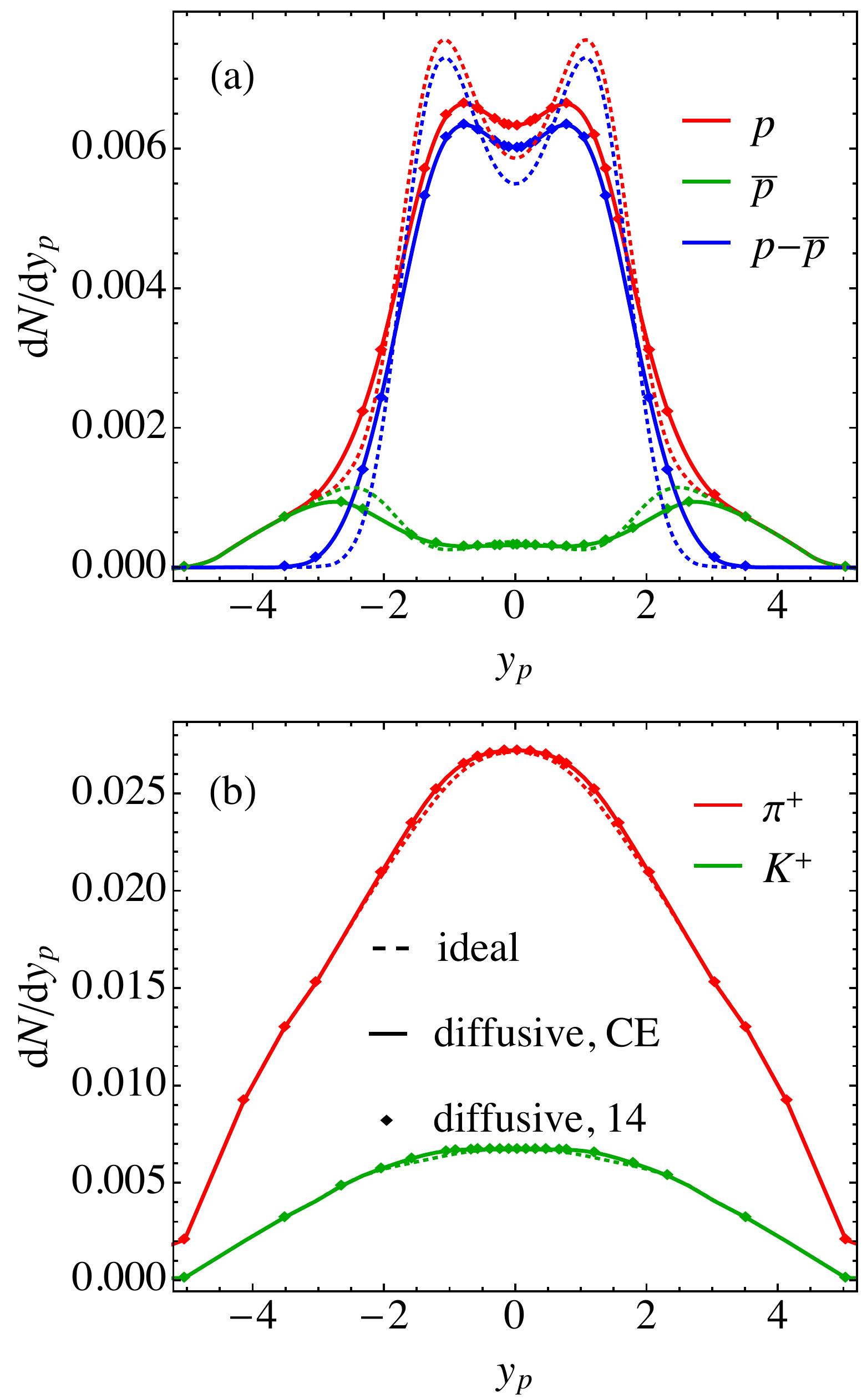}
\caption{%
    Final particle rapidity distributions for (a) baryons (red: protons; green: antiprotons; blue: net protons) and (b) mesons (red: pions; green: kaons). Thin dashed and thick solid lines show results for ideal and diffusive evolution, respectively. For the diffusive case, the solid lines use    the Chapman-Enskog approximation for the diffusive correction, while colored markers show the values obtained from the 14-moment approximation.}
    \label{fig:1D_spec_comp}
\end{center}
\end{figure}

Given these quantities on the freeze-out surface, we use the \isd{} module \cite{McNelis:2019auj} to evaluate the Cooper-Frye integral (\ref{eq:CooperFrye},\ref{eq:diffusion_deltaf}) for the rapidity distributions of hadrons emitted from the freeze-out surface. Results are shown in Fig.~\ref{fig:1D_spec_comp}. Panel (b) indicates that baryon diffusion has negligible effects on meson distributions. It affects only baryon distributions. Panel (a) shows that baryon diffusion significantly increases the proton and net-proton yields at mid-rapidity and also broadens their rapidity distributions at large rapidity. \ld{Both of these effects on the baryon distributions were also observed in earlier work that, different from our simulations, additionally included resonance decays and full hadronic rescattering \cite{Shen:2017ruz, Du:2018mpf, Denicol:2018wdp, Li:2018fow}; furthermore, they were found to increase with the magnitude of the baryon diffusion coefficient $\kappa_n$.} The approximate boost-invariance of the longitudinal flow over a wide range of \etas{} on the freeze-out surface (see Fig.~\ref{fig:fz_surf}b) maps the baryon diffusion effects seen in Figs.~\ref{fig:long_evolution}d,e and \ref{fig:fz_surf}c as functions of space-time rapidity $\eta_s$ onto momentum rapidity $y_p$ in Fig.~\ref{fig:1D_spec_comp}a \cite{Shen:2017ruz, Du:2018mpf, Denicol:2018wdp, Li:2018fow}. \ld{We take advantage of the \isd{} option to include both the Chapman-Enskog and 14-moment approximations for the dissipative corrections (\ref{eq:diffusion_deltaf}), comparing the two in Fig.~\ref{fig:1D_spec_comp}. The difference is seen to be negligibly small, and even ignoring in Eq.~(\ref{eq:CooperFrye}) $\delta f_{\mathrm{diff},i}$ entirely does not make much of a difference (not shown).} This reflects the tiny magnitude of the baryon diffusion current on the freeze-out surface seen in Fig. \ref{fig:fz_surf}d.\footnote{%
    Ref.~\cite{Denicol:2018wdp}, with transverse expansion, shows that baryonic observables in the transverse plane, such as the $p_T$-differential proton elliptic flow $v_2^p(p_T)$, are sensitive to the dissipative correction $\delta f_\mathrm{diff}$ from baryon diffusion.}

We emphasize that the mapping of baryon diffusion effects seen as a function of spacetime rapidity \etas{} in Figs.~\ref{fig:long_evolution}d,e and \ref{fig:fz_surf}c onto momentum rapidity $y_p$ is expected to be model dependent, and may not work for initial conditions in which the initial velocity profile is not boost-invariant or the initial \etas{}-distribution of the net baryon density looks different. This initial-state modeling uncertainty has so far prohibited a meaningful extraction of the baryon diffusion coefficient from experimental data (see, however, Ref.~\cite{Denicol:2018wdp} for a valiant effort). Additional uncertainties from possible critical effects associated with QCD critical point on the bulk dynamics, especially through baryon diffusion, may further complicate the picture, in particular as long as the location of the critical point is still unknown. In the following subsection we address some of these effects arising from critical dynamics.

\subsection{Critical effects on baryon diffusion}
\label{sec:diffcp}

In this section, we explore whether the QCD critical point can have significant effects on the bulk dynamics, through the baryon diffusion current. For this purpose, we include critical effects as described in Sec.~\ref{sec:critical}, and explore effects from critical slowing down on the hydrodynamic transport (Sec.~\ref{sec:diffcp_hydro}), as well as critical corrections to final particle distributions through the Cooper-Frye formula (Sec.~\ref{sec:diffcp_spectra}).

\subsubsection{Critical slowing down of baryon transport}
\label{sec:diffcp_hydro}

As discussed in Sec.~\ref{sec:critical}, in the critical region
baryon transport is affected by critical slowing down \cite{RevModPhys.49.435}. Outside the critical region all thermodynamic and transport properties approach their non-critical baseline described in Sec.~\ref{sec:diffnocp}, but as the system approaches the critical point its dynamics is affected by critical modifications of the transport coefficients involving various powers of $\xi/\xi_0>1$. We study this by incorporating the critical scaling of $\chi$, $\kappa_n$ and $\tau_n$ in Eqs.~\eqref{eq:chi_kappa_scaling} and \eqref{eq:tau_n_xi}, with the correlation length $\xi(\mu,T)$ parametrized by Eq.~\eqref{eq:xi(mu,T)}. 

Before doing any simulations we briefly discuss qualitative expectations. Eq.~\eqref{eq:D_scaling} indicates that, as the correlation length grows, $\xi/\xi_0>1$, the coefficient $D_B$ is suppressed while $D_T$ is enhanced. According to  Eqs.~\eqref{eq:nmu_NS_decomposition} a suppression of $D_B$ reduces the contribution from baryon density inhomogeneities while an enhancement of $D_T$ increases the contribution from  temperature inhomogeneities to the Navier-Stokes limit $n^\nu_\mathrm{NS}=\kappa_n\nabla^\nu(\mu/T)$.\footnote{%
    We note that in the literature sometimes only the baryon density gradient term $D_B\nabla n$ is included in the diffusion current (see, e.g., Refs.~\cite{RevModPhys.49.435, Son:2004iv}) which then leads to its generic suppression close to the critical point.}
In addition to thus moving its Navier-Stokes target value, proximity of the critical point also increases the time $\tau_n$ (see Eq.~\eqref{eq:tau_n_xi}) over which the baryon diffusion current relaxes to its Navier-Stokes limit -- its response to the driving force is critically slowed down. 

\begin{figure}[!tbp]
\begin{center}
\hspace{-0.5cm}
\includegraphics[width= 0.4\textwidth]%
    {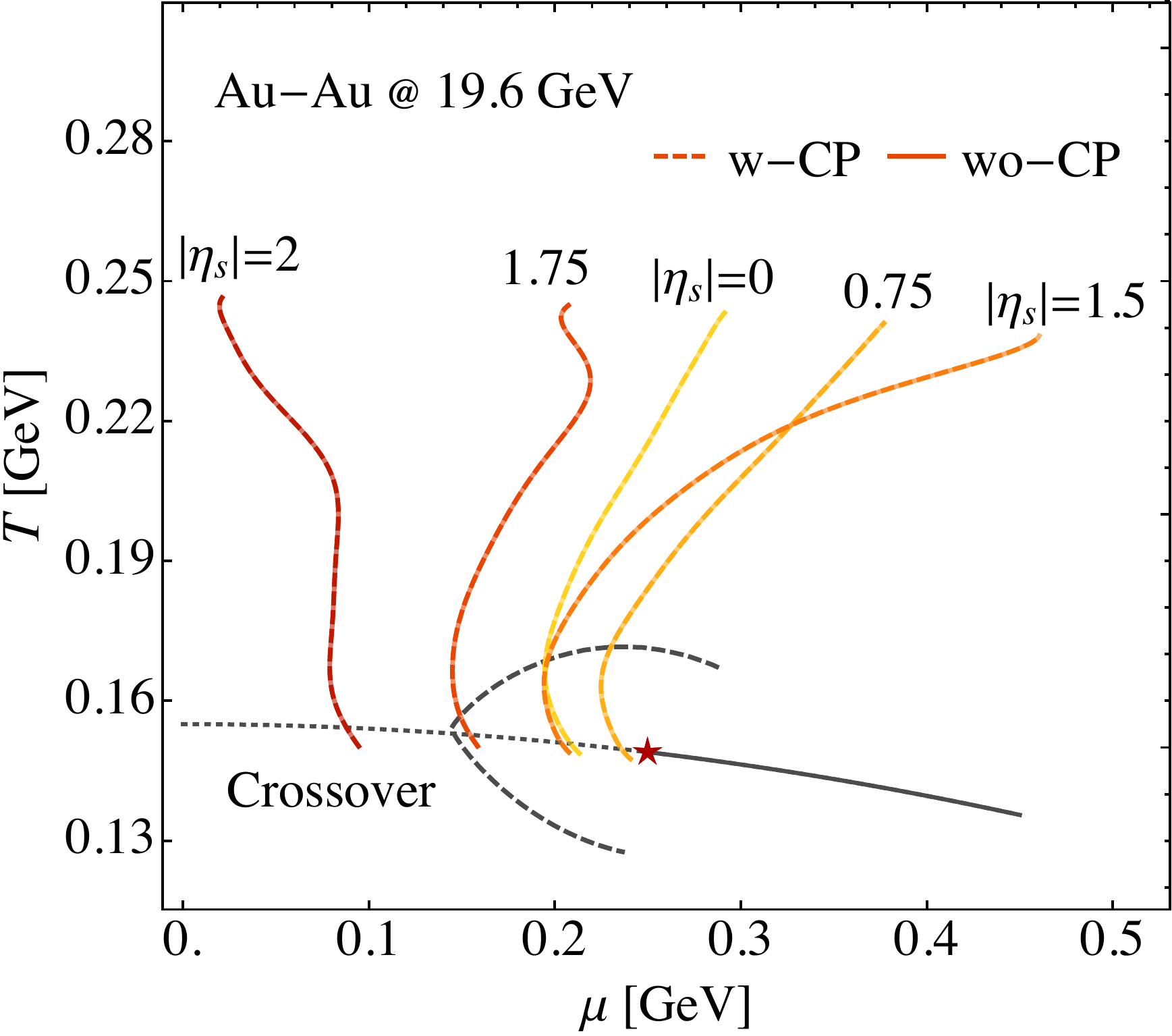}
\caption{%
    Phase diagram trajectories at different space-time rapidities, with (w-CP, dashed lines) and without (wo-CP, solid lines) inclusion of critical effects. The solid lines are taken from Fig.~\ref{fig:phase_dia_traj}. The dashed lines account for the critical scaling of all parameters controlling the evolution. 
    The black dashed line encloses the region on the crossover side where $\xi(\mu,T)>\xi_0\eq1$\,fm.}
    \label{fig:trajectory_CP_noCP}
\end{center}
\end{figure}

Repeating the simulations with the same setup as in Sec.~\ref{sec:diffnocp}, except for the inclusion of critical scaling, yields the results shown in Fig.~\ref{fig:trajectory_CP_noCP}. For the parametrization of the correlation length $\xi(\mu,T)$ we assumed a critical point located at ($T_c\eq149$\,MeV,\,$\mu_c\eq250$\,MeV). This is very close to the right-most trajectory shown in Fig.~\ref{fig:trajectory_CP_noCP} which should therefore be most strongly affected by it.\footnote{%
    Since we do not have the tools here to handle passage through a first-order phase transition, we do not consider any expansion trajectories cutting the first-order transition line to the right of QCD critical point.}
Surprisingly, none of the trajectories, not even the one passing the critical point in close proximity, are visibly affected by critical scaling of transport coefficients.

\begin{figure}[!tbp]
\begin{center}
\hspace{-0.5cm}
\includegraphics[width=0.36\textwidth]%
                {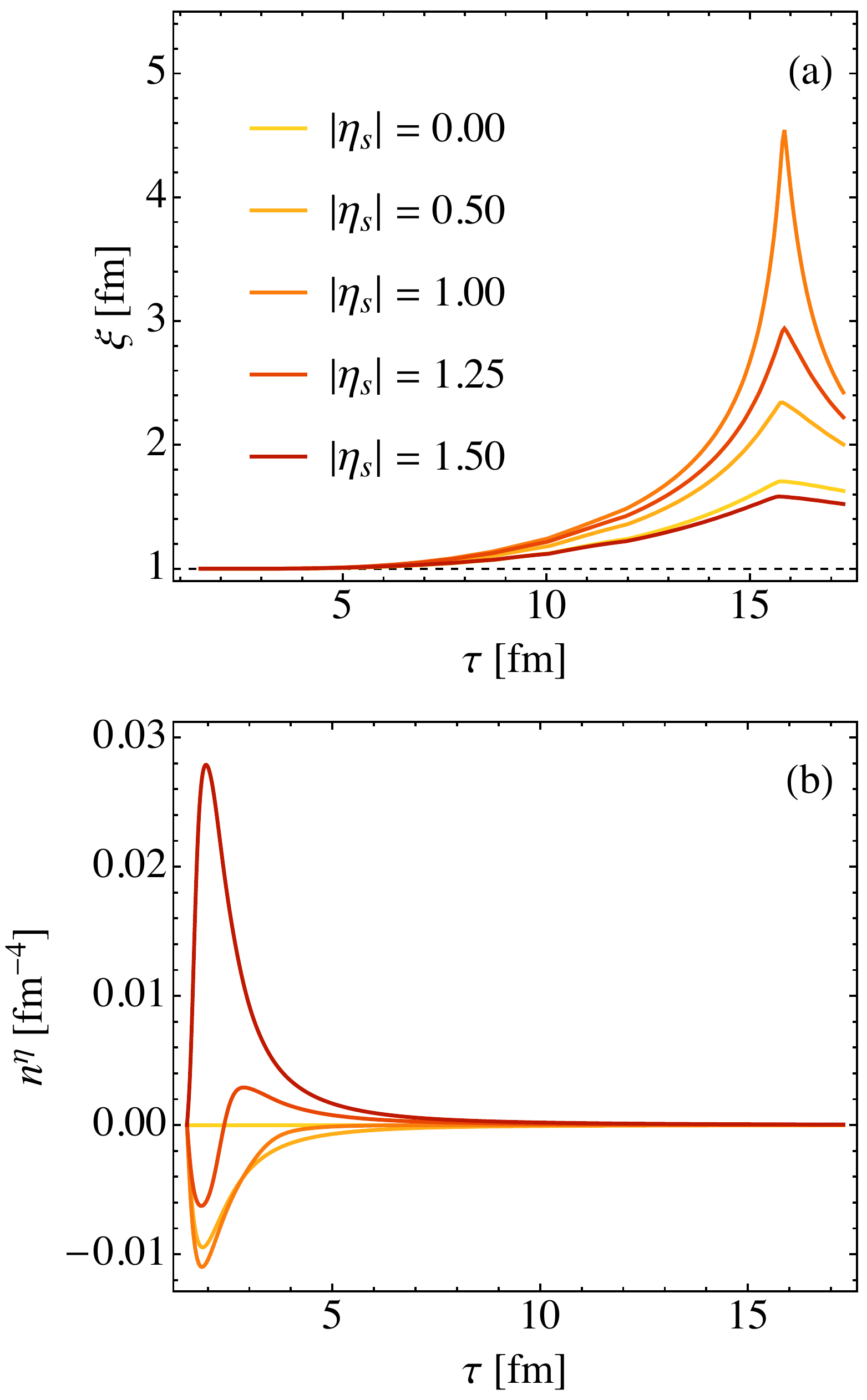}
\caption{%
    Time evolution of (a) correlation length and (b) longitudinal baryon diffusion at selected space-time rapidities. Note that $n^\eta$ at $\eta_s=1.25$ is shown intentionally as its sign changes during the evolution. See text for discussion.
    }
    \label{fig:xi_neta_evolution}
\end{center}
\end{figure}

To better understand this we plot in Fig.~\ref{fig:xi_neta_evolution} the history of the correlation length and baryon diffusion current at different $\eta_s$. In panel (a) we see that $\xi$ does show the expected critical enhancement, by up to a factor $\sim 4.5$ at $\eta_s\eq1$. This maximal enhancement corresponds to $\tau_n\simeq20\,\tau_{n,0}$ and $D_B\simeq0.22\,D_{B,0}$, naively suggesting significant effects on the dynamical evolution. However, the critical enhancement of the correlation length does not begin in earnest before the fireball has cooled down to a low temperature $T\lesssim T_c+\Delta T$. 
%
%
Fig.~\ref{fig:xi_neta_evolution}b shows at at this late time the baryon diffusion current has already decayed to a tiny value.\footnote{%
    This statement remains true if one multiplies $n^\eta$ with the metric factor $\sqrt{-g}=\tau$ to obtain the baryon diffusion current in physical units of fm$^{-3}$.}
In other words, the largest baryon diffusion currents are created at early times when the temporal gradients are highest but the system is far from the critical point; by the time the system gets close to the critical point, thermal and chemical gradients have decayed to such an extent that even a critical enhancement of the correlation length by a factor 5 can no longer revive the baryon diffusion current to a noticeable level. 

\begin{figure}[!tbp]
\begin{center}
\hspace{-0.5cm}
\includegraphics[width= 0.36\textwidth]{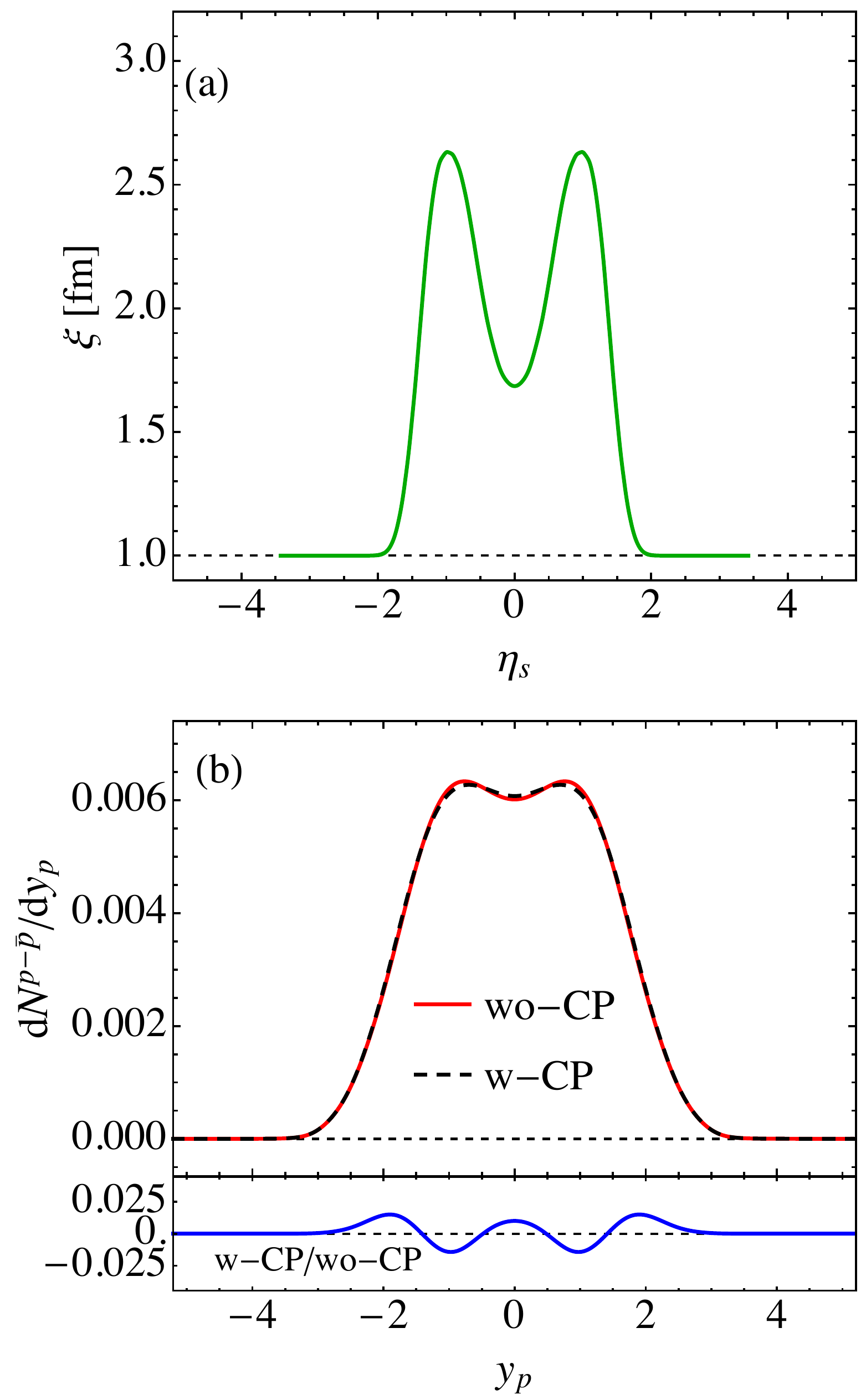}
\caption{%
    (a) Space-time rapidity distribution of the correlation length along the freeze-out surface. (b) {\sl Upper panel:} Net proton rapidity distributions with and without critical scaling effects (w-CP and wo-CP), calculated from the Cooper-Frye formula \eqref{eq:CooperFrye} with diffusive correction \eqref{eq:diffusion_deltaf}. For the w-CP case critical scaling effects are included in both the hydrodynamic evolution and the diffusive correction at particlization. {\sl Lower panel:} Deviation from 1 of the ratio of the rapidity spectra shown in the upper panel.}
    \label{fig:xi_spec_cp}
\end{center}
\end{figure}

This two-stage feature, with a first stage characterized by large baryon diffusion effects without critical modifications and a second stage characterized by large critical  fluctuations \cite{Rajagopal:2019xwg, Du:2020bxp} with negligible baryon diffusion effects on the bulk evolution, is an important observation. For a deeper understanding we devote Sec.~\ref{sec:tevol} to a more systematic investigation of the time evolution of the diffusion current, but not before a brief exploration in the following subsection of critical effects on the final single-particle distributions.

\subsubsection{Critical corrections to final single-particle distributions}
\label{sec:diffcp_spectra}

How to consistently include critical fluctuation effects on the finally emitted single-particle distributions, at the ensemble-averaged level, is still a subject of active research. A solid framework may require a microscopic picture involving interactions between the underlying degrees of freedom and the fluctuating critical modes during hadronization \cite{Stephanov_2010}. In this work we employ a simple ansatz where critical corrections to the final particle distributions are included only via the diffusive correction $\delta f_\mathrm{diff}$ from Eq.~\eqref{eq:diffusion_deltaf} appearing in the Cooper-Frye formula \eqref{eq:CooperFrye}. In this subsection this dissipative correction is computed from the simulations described in the preceding subsection which include critical correlation effects through critically modified transport coefficients, specifically a normalized baryon diffusion coefficient $\hat{\kappa}\equiv\kappa_n/\tau_n$ with critical scaling
\begin{equation}
\label{eq:critical_deltaf}
    \hat{\kappa}=\hat{\kappa}_0\left(\frac{\xi}{\xi_0}\right)^{-1}\,,
\end{equation}
obtained from Eqs.~\eqref{eq:chi_kappa_scaling} and \eqref{eq:tau_n_xi} using $\hat\kappa_0 = \kappa_{n,0}/ \tau_{n,0}$.\footnote{%
    We note that this ansatz is more straightforwardly implemented in the Chapman-Enskog method used here than in the 14-moment approximation \cite{Monnai:2018rgs,McNelis:2019auj} whose  transport coefficients do not have such obvious critical scaling.}
Since we saw in the preceding subsection that the hydrodynamic quantities on the particlization surface are hardly affected by the inclusion of critical scaling effects during the preceding dynamical evolution, the main critical scaling effects on the emitted particle spectra arise from any critical modification that $\hat{\kappa}$ might experience on the particlization surface. 

The space-time rapidity distribution of the correlation length $\xi$ along the freeze-out surface, as well as the net proton rapidity distributions with and without critical scaling effects, are shown in Fig.~\ref{fig:xi_spec_cp}. Panel (a) shows that $\xi$ peaks near $|\eta_s|\simeq1.0$ on the freeze-out surface, consistent with Fig.~\ref{fig:xi_neta_evolution}a. Note that, although fluid cells at different $\eta_s$ generally freeze out at different times, the freeze-out surface in Fig.~\ref{fig:fz_surf}a shows that within $\eta_s\in[-1.5, 1.5]$ all fluid cells freeze out at basically the same time $\tau_f\sim17$\,fm/$c$. Therefore Fig.~\ref{fig:xi_spec_cp}a indeed corresponds to the $\xi$ values at different \etas{} at the end of the evolution in Fig.~\ref{fig:xi_neta_evolution}a.

Even though Fig.~\ref{fig:xi_spec_cp}a shows a critical enhancement of $\xi/\xi_0{\,\simeq\,}2.7$ near $\eta_s{\,\simeq\,}1.0$, corresponding to $\hat{\kappa}/\hat{\kappa}_0{\,\simeq\,}0.37$, we see in Fig.~\ref{fig:xi_spec_cp}b that the net proton distribution is modified by at most a few percent. The lower panel in Fig.~\ref{fig:xi_spec_cp}b indicates that the largest critical corrections indeed correspond to regions of large $\xi/\xi_0$, sign-modulated by the direction of the baryon diffusion current (cf. Fig.~\ref{fig:fz_surf}d). We also notice a thermal smearing when mapping the distribution of $\xi$ in \etas{} to the modification of net proton distribution in $y_p$. The critical modification of the net proton spectra arising from the diffusive correction to the distribution function is very small also because $\delta f_\mathrm{diff}$ in \eqref{eq:diffusion_deltaf} is roughly proportional to the magnitude of $n^\mu$ which is tiny. Such small modifications are certainly unresolvable with current or expected future measurement. \ld{Indeed, the magnitudes of these modifications depend on the correlation length on the freeze-out surface which in turn depends on the choice of the freeze-out energy density $e_\mathrm{f}$. Independence of the final results from this choice could be achieved by properly sampling the critical correlations on this surface and then propagating them to the completion of kinetic freeze-out with a hadronic transport code that appropriately accounts for critical dynamics; unfortunately, these options are presently not yet available.}  

In conclusion, critical scaling effects on both the hydrodynamic evolution of the bulk medium and the finally emitted single-particle momentum distributions are small, mostly because by the time the system passes the critical point and freezes out the baryon diffusion current has decayed to negligible levels.

\subsection{Time evolution of baryon diffusion}
\label{sec:tevol}

In this subsection we further analyze the baryon diffusion dynamics and the origins of its rapid decay. We define Knudsen and inverse Reynolds numbers for baryon diffusion and display their space-time dynamics. The resulting insights are relevant for model building and for the future quantitative calibration of the bulk fireball dynamics at non-zero chemical potential. 

\subsubsection{Fast decay of baryon diffusion}
\label{sec5d1}

As discussed in Sec.~\ref{sec:hydro}, the diffusion current relaxes to its Navier-Stokes limit $n^\nu_\mathrm{NS}\eq\kappa_n\nabla^\nu(\mu/T)$ on a time scale given by $\tau_n$. General features of baryon diffusion evolution can thus be understood by following the time evolution of $n^\mu_\mathrm{NS}$, $\kappa_n$ and $\tau_n$. Here we focus on their evolution without inclusion of critical scaling since we established that the latter has negligible effect on the bulk evolution and therefore the non-critical values of $n^\mu_\mathrm{NS}$, $\kappa_n$ and $\tau_n$ evolve almost identically with and without inclusion of critical effects.

\begin{figure}[!tbp]
\begin{center}
\hspace{-0.5cm}
\includegraphics[width= 0.4\textwidth]{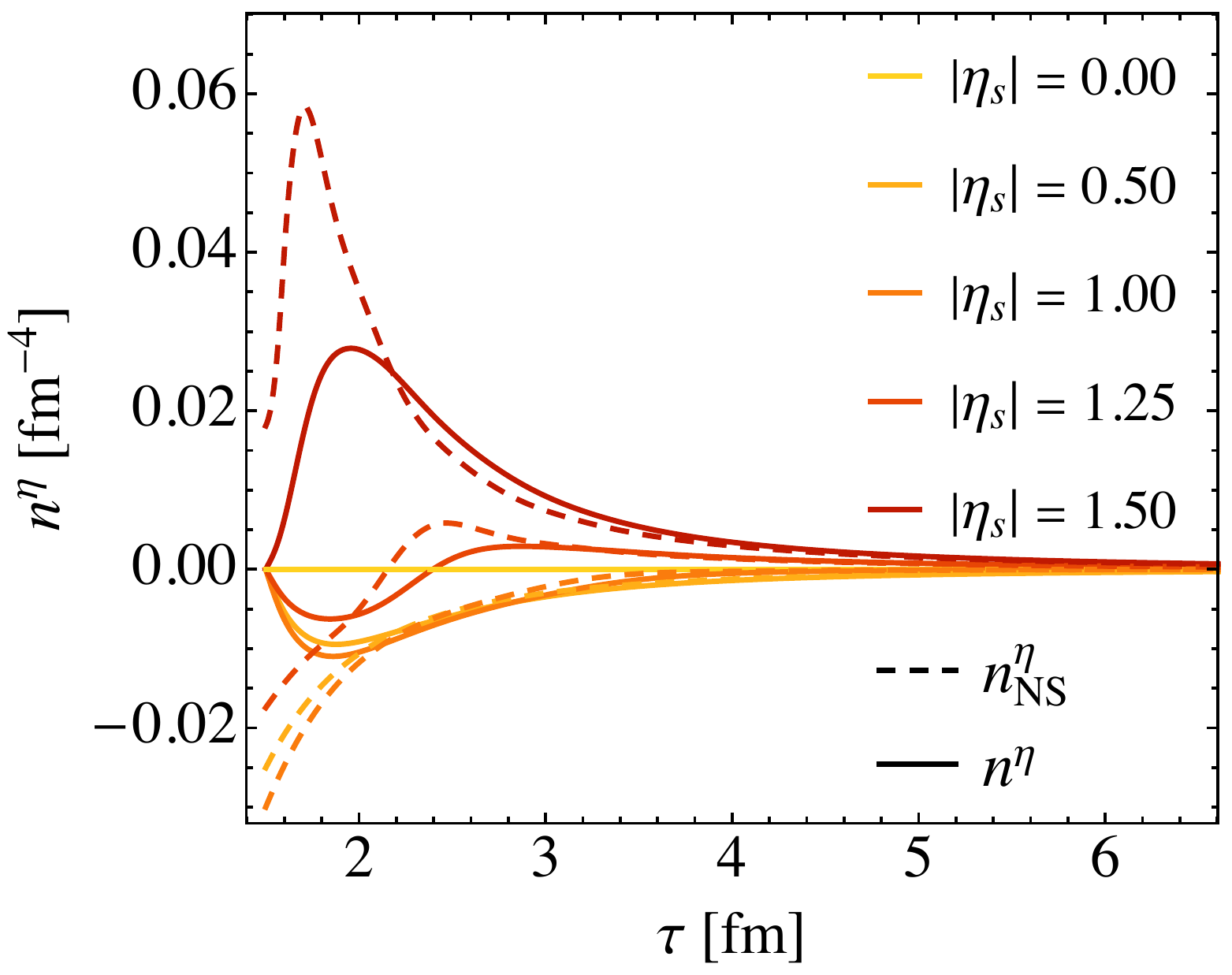}
\caption{%
    Same as Fig.~\ref{fig:xi_neta_evolution}b, but zoomed in onto the early evolution stage and including for comparison as dashed lines the corresponding Navier-Stokes limit $n^\eta_\mathrm{NS}$ of the longitudinal diffusion current.
    \label{fig:NSlimit_evolution}}
\end{center}
\end{figure}

\begin{figure}[!tbp]
\begin{center}
    \hspace{-0.5cm}\includegraphics[width= 0.36\textwidth]{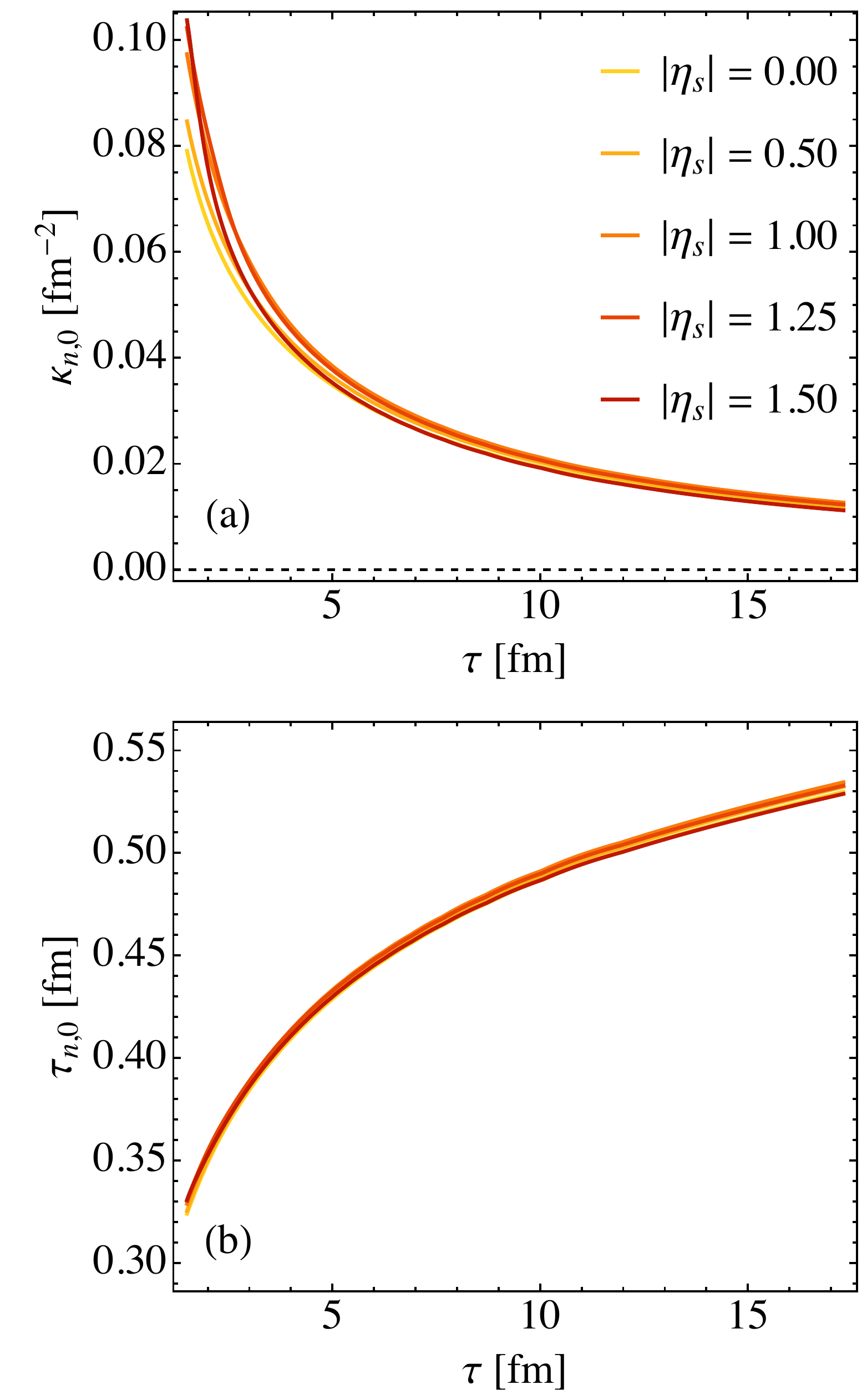}
    \caption{Time evolution of (a) $\kappa_{n,0}$ and (b) $\tau_{n,0}$ at some space-time rapidities, corresponding to Figs.~\ref{fig:xi_neta_evolution} and \ref{fig:NSlimit_evolution}.}
    \label{fig:kappa_tau_evolution}
\end{center}
\end{figure}

Fig.~\ref{fig:NSlimit_evolution} shows a comparison of the longitudinal baryon diffusion current (solid lines) with its Navier-Stokes limit (dashed lines) at different space-time rapidities. One sees that the relaxation equation for $n^\eta$ tries to align the diffusion current with its Navier-Stokes value (which is controlled by the longitudinal gradient $\nabla^\eta(\mu/T)$) but the finite relaxation time delays the response, causing $n^\eta$ to perform damped oscillations around $n^\eta_\mathrm{NS}$. This is most clearly illustrated in Fig.~\ref{fig:NSlimit_evolution} by following the cell located at $\eta_s\eq1.5$ (uppermost): Initialized at zero, $n^\eta$ initially rises steeply, trying to adjust to its positive and rapidly increasing Navier-Stokes value, but at $\tau{\,\simeq\,}1.7$\,fm/$c$ the longitudinal gradient of $\mu/T$ switches sign and $n^\eta_\mathrm{NS}$ starts to decrease again. The hydrodynamically evolving $n^\eta$ follows suit, turning downward with a delay of about 0.3\,fm/$c$ (which, according to Fig.~\ref{fig:kappa_tau_evolution}b below, is the approximate value of the relaxation time $\tau_{n,0}$ at $\tau\eq2$\,fm/$c$), but soon finds itself overshooting its Navier-Stokes value. For the cell located at $\eta_s\eq1.25$, $n^\eta$ crosses its Navier-Stokes value even twice. 

As long as the relaxation time $\tau_n$ is short and not dramatically increased by critical slowing down, the rapid decrease of the dynamically evolving diffusion current is seen to be a generic consequence of a corresponding rapid decrease of its Navier-Stokes value: Fig.~\ref{fig:NSlimit_evolution} shows that after $\tau{\,\sim\,}3.5$\,fm/$c$, $n^\eta$ basically agrees with its Navier-Stokes limit $n^\eta_\mathrm{NS}$. Fig.~\ref{fig:kappa_tau_evolution}b shows that in the absence of critical effects the relaxation time $\tau_{n,0}\eq C_n/T$ increases by less than a factor of 2 over the entire fireball lifetime. The rapid decrease of $n^\eta_\mathrm{NS}$ is a consequence of two factors: (i) the gradients of $\mu/T$ decrease with time, owing to both the overall expansion of the system and the diffusive transport of baryon charge from dense to dilute regions of net baryon density, and (ii) the baryon diffusion coefficient $\kappa_{n,0}$ decreases dramatically (by almost an order of magnitude over the lifetime of the fireball as seen in Fig.~\ref{fig:kappa_tau_evolution}a), as a result of the fireball's decreasing temperature.

In summary, three factors contribute to the negligible influence of the QCD critical point on baryon diffusion: First, baryon diffusion is largest at very early times when its relaxation time is shortest and it quickly relaxes to its Navier-Stokes value; the latter decays quickly, due to decreasing chemical gradients and a rapidly decreasing baryon diffusion coefficient. Second, the relaxation time for baryon diffusion increases at late times, generically as a result of cooling but possibly further enhanced by critical slowing down if the system passes close to the critical point. This makes it difficult for the baryon diffusion current to grow again. Third, critical effects that would modify\footnote{%
    According to Eq.~\eqref{eq:nmu_NS_decomposition} and Eq.~(\ref{eq:D_scaling}), critical effects can increase or decrease the Navier-Stokes value of the baryon diffusion, depending on the relative sign and magnitude of the density and temperature gradients.}
the Navier-Stokes limit for the baryon diffusion current become effective only at very late times when $n^\eta_\mathrm{NS}$ has already decayed to non-detectable levels. The baryon diffusion current thus remains small even if its Navier-Stokes value were significantly enhanced by critical scaling effects.

\subsubsection{Knudsen and inverse Reynolds numbers}

\begin{figure}[!tbp]
\begin{center}
\hspace{-.8cm}
\includegraphics[width= 0.52\textwidth]{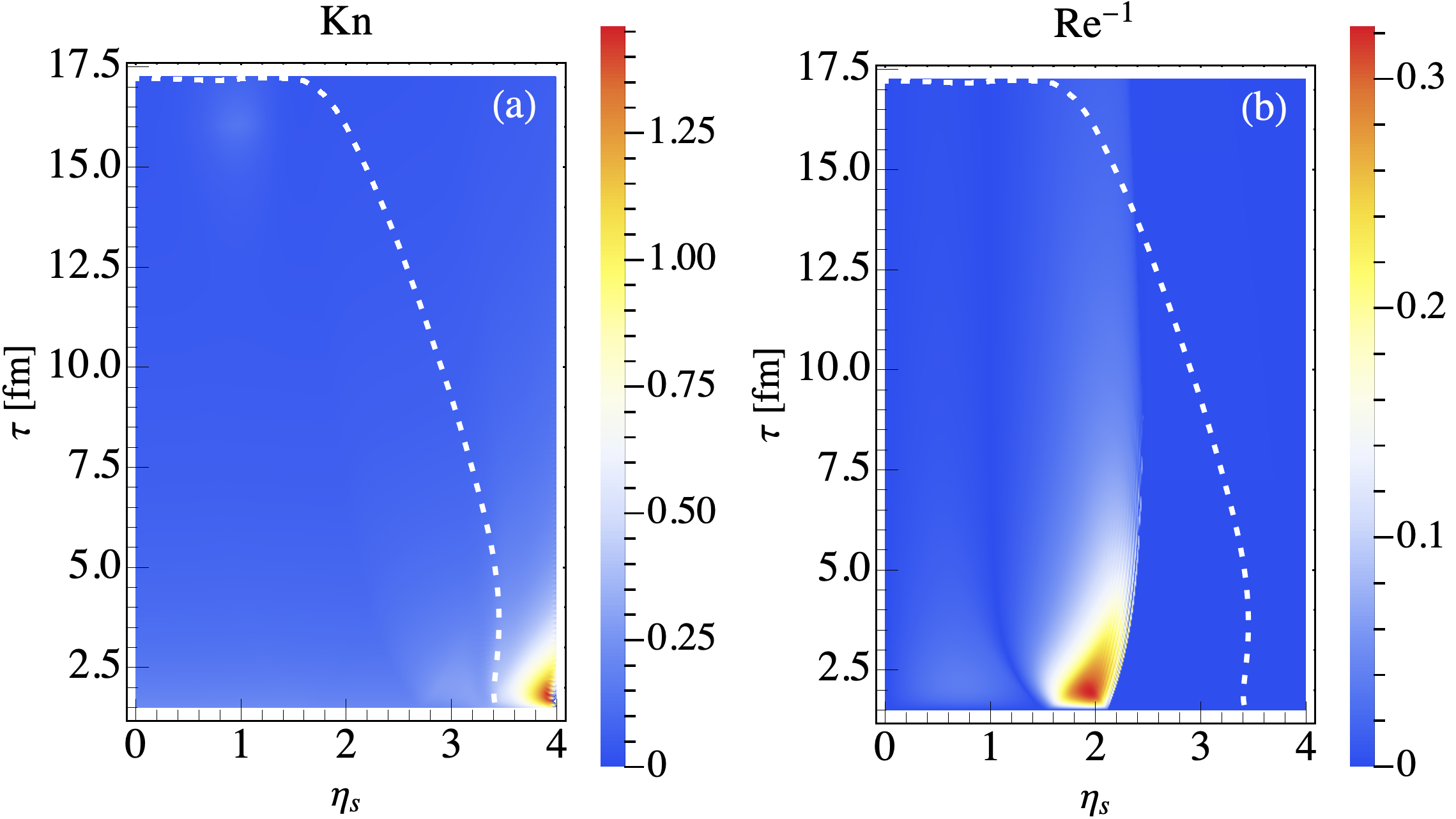}
\caption{%
    Space-time evolution of (a) the Knudsen number Kn and (b) the inverse Reynolds number $\mathrm{Re}^{-1}$ for the baryon diffusion current. The white dashed line shows the same freeze-out surface as Fig.~\ref{fig:fz_surf}a.
    \label{fig:kn_re_evolution}}
\end{center}
\end{figure}

We close this section by investigating the (critical) Knudsen and inverse Reynolds numbers associated with baryon diffusion. These are typically taken as quantitative measures to assess the applicability of second order viscous hydrodynamics such as the \beshydro{} framework employed in this work. Copying their standard definitions for shear and bulk viscous effects \cite{Shen:2014vra, Denicol:2018wdp, Du:2019obx}, we here set
\begin{equation}
    \mathrm{Kn}\equiv \tau_n\theta\,,\quad\mathrm{Re}^{-1}\equiv\frac{\sqrt{|n^\mu n_\mu|}}{n}\,,
\end{equation}
for baryon diffusion, where $\theta$ is the scalar expansion rate. Kn is the ratio between time scales for microscopic diffusive relaxation ($\tau_n$) and macrosopic expansion ($\tau_\mathrm{exp}=1/\theta$); the relaxation time $\tau_n$ includes the effects of critical slowing down in the neighborhood of the QCD critical point. $\mathrm{Re}^{-1}$ is the ratio between the magnitude of the off-equilibrium baryon diffusion current and the equilibrium net baryon density in ideal fluid dynamics. Their space-time evolutions are shown in Fig.~\ref{fig:kn_re_evolution}, together with the freeze-out (particlization) surface at $e_\mathrm{f}\eq0.3\,$GeV/fm$^3$.

Fig.~\ref{fig:kn_re_evolution}a tells us that $\mathrm{Kn}{\,\gtrsim\,}1$ happens only outside the freeze-out surface, in the fireball's corona where the fluid has already broken up into particles even at the earliest stage of the expansion. The short-lived peak in Kn near $\tau\eq\tau_i=1.5$\,fm/$c$ and $\eta_s{\,\sim\,}4$ is caused by the rapid increase of $\tau_n$ in the dilute and very cold corona of the fireball (note that $\tau_{n,0}=C_n/T$). Critical slowing down near the QCD critical point causes the Knudsen number to increase somewhat around $\eta_s\eq1$ close to the freeze-out surface; this critical enhancement is barely visible as a light cloud on a blue background, indicating critical Knudsen numbers in the range Kn${\,\sim\,}0.5-0.7$. Fig.~\ref{fig:kn_re_evolution}b, on the other hand,  indicates that $\mathrm{Re}^{-1}{\,\lesssim\,}0.3$ during the entire evolution, even close to the places where the Navier-Stokes value of the baryon diffusion current peaks at early times (see Fig.~\ref{fig:NSlimit_evolution}). After $\tau{\,\sim\,}5$\,fm/$c$ (including the entire critical region around the QCD critical point) its maximum value drops below 0.1, reflecting of the rapid decay of the baryon diffusion current. 

The maximal $\mathrm{Re}^{-1}$ occurs around $\eta_s\simeq2$ shortly after the hydrodynamic evolution starts at $\tau_i=1.5\,$fm/$c$. From it emerges a region of sizeable inverse Reynolds number which ends at two moving boundaries where $\mathrm{Re}^{-1}\eq0$ (dark blue). The left boundary, moving towards smaller \etas{} values, reflects a sign change of the baryon diffusion current (see Fig.~\ref{fig:long_evolution}f where at $\tau\eq5.5$\,fm/$c$ $n^\eta$ flips sign at $\eta_s\simeq1$). The right boundary, on the other hand, corresponds to where $n^\mu$ decays to zero (which, according to Fig.~\ref{fig:long_evolution}f, happens at $\eta_s{\,\simeq2.5\,}$ when $\tau\eq5.5$\,fm/$c$). The initial outward movement of the right boundary is a result of baryon transport to larger space-time rapidity. It stops moving after the diffusion current has decayed and no longer transports any baryon charge longitudinally.

The small values of $\mathrm{Kn}$ and $\mathrm{Re}^{-1}$ during the entire fluid dynamical evolution validate the applicability of second order viscous hydrodynamics, \beshydro, for describing flow and diffusive transport of baryon charge in the collision system studied here.

\section{Conclusions and discussion}
\label{sec:summary}

Baryon diffusion is an important dissipative effect in the hydrodynamic evolution of systems carrying a conserved net baryon charge. It smoothes out chemical inhomogeneities by transporting baryon charge relative to the momentum flow from regions of large to smaller net baryon density. It is driven by gradients of $\mu/T$, but the corresponding transport coefficient characterizing the magnitude of the diffusive response, the baryon diffusion coefficient $\kappa_n(\mu,T)$, is still poorly constrained theoretically. In this work we studied a (1+1)-dimensional system without transverse gradients and flow to explore diffusive baryon transport along the longitudinal (beam) direction in heavy-ion collisions and thereby gain intuition about possible strategies to extract the baryon diffusion coefficient from experimental data.

A fundamental difficulty for such an extraction is that baryon diffusion manifests itself as a transport of net baryon density {\em in coordinate space} while the experimental observations provide a snapshot of the hydrodynamic medium at the end of its lifetime {\em in momentum space}. In a rapidly expanding system, collective flow gradients map different space-time regions into different regions of momentum space, but the thermal momentum spread in the local rest frame blurs this map (less so for the heavier baryons than for the more abundant lighter mesons), and makes it difficult to reconstruct the movement generated by baryon diffusion in coordinate space from the final net baryon distribution in momentum space. In longitudinal direction local thermal motion results in a rapidity spread of order $\sqrt{T/\langle m_T\rangle}$ \cite{Schnedermann:1993ws, Chapman:1994ax} where $T$ is the kinetic freeze-out temperature and $\langle m_T\rangle$ is the average transverse mass of the particle species in question. The flow-induced separation of different fireball regions in longitudinal momentum rapidity is easier at higher collision energies where the fluid medium created in the collision covers a wider rapidity range, i.e. a larger multiple of the thermal smearing width. At lower beam energies, such as those probed in the RHIC BES campaign, unfolding the measured final rapidity distribution into different regions of space-time rapidity, with possibly different net baryon densities, becomes much harder. And therefore the reconstruction of baryon-diffusion induced matter transport across space-time rapidity also becomes much harder.

In this work we studied central Au-Au collisions at $\sqrt{s_\mathrm{NN}}\sim20$\,GeV in which the fireball covers about 7 units of space-time rapidity along the beam direction (Fig.~\ref{fig:long_evolution}a). Assuming that baryon stopping leads to a space-time rapidity shift of about 1.5 units for the incoming projectile and target baryons, the initial net baryon distribution had a width of about 4 units. It was modeled by a double-humped function with two well-separated peaks located at $\eta_s\eq\pm1.5$ (Fig.~\ref{fig:long_evolution}d). After accounting for ideal hydrodynamic evolution and thermal smearing this resulted in a double-humped net proton rapidity distribution whose peaks were still relatively cleanly separated by about 2 units of rapidity, but after including baryon diffusion they almost (though not quite) merged into a single broad hump around midrapidity (Fig.~\ref{fig:1D_spec_comp}a). In our calculation, the QCD critical point was positioned at a baryon chemical potential $\mu=250$\,MeV. Recent Lattice QCD results put the likely location of this critical point at $\mu>400$\,MeV \cite{Bazavov:2017dus, Mukherjee:2019eou, Giordano:2019gev}, which requires lower collision energies for its experimental exploration. At lower collision energies the width of the initial space-time rapidity interval of non-zero net baryon density will be narrower, and the final net-proton distribution will eventually become single-peaked in central collisions \cite{Bearden:2003hx}.   

In the work presented here we focused on the questions how diffusive baryon transport manifests itself along the beam direction in hydrodynamic simulations, what traces it leaves in the finally measured rapidity distributions, and how it is affected by critical scaling of transport coefficients in the proximity of the QCD critical point. To address these questions we systematically discussed the static and dynamic critical behavior of thermodynamic properties (especially those associated with baryon transport) and introduced an analytical parametrization of the correlation length that correctly reproduces the critical exponents of the 3D Ising universality class. Based on a careful comparison with the Hydro+/++ framework \cite{Stephanov:2017ghc, An:2019hydro++, Monnai:2016kud} we identified the critical scaling (``critical slowing down'') of the relaxation time for the baryon diffusion current ($\tau_n\sim\xi^2$), and demonstrated that in the critical regime the Israel-Stewart type equation for the baryon diffusion current plays the role of a single-mode Hydro+ equation for a vector slow mode. We did not discuss the out-of-equilibrium evolution of the slow mode itself. For a single scalar slow mode, this was studied elsewhere using the \beshydro+ framework \cite{Du:2020bxp} where it was found that its feedback to the hydrodynamic bulk evolution was negligible \cite{Du:2020bxp,Rajagopal:2019xwg}. A systematic Hydro++ study incorporating the full coupled evolution of all relevant non-hydrodynamic critical slow modes is still outstanding.

We are not the first to point out that, because of its extended nature, different regions within a collision fireball probe different regions of the QCD phase diagram as they cool and expand. We found that, at early times, strong longitudinal gradients of $\mu/T$ lead to significant longitudinal baryon diffusion currents that shift the expansion trajectories for different parts of the fireball in different directions within the QCD phase diagram. The final net proton rapidity distributions reflect these shifts, albeit blurred by thermal smearing. Our model assumes a boost-invariant initial longitudinal momentum flow $y_\mathrm{flow}=\eta_s$ which maximizes the correlation between the space-time rapidity of a fluid cell on the freeze-out surface and the momentum rapidity of the final hadrons it emits. The expected breaking of boost-invariance of the longitudinal flow pattern by dynamical initialization effects at lower collision energies even near midrapidity \cite{Shen:2017bsr,Du:2018mpf,du2020ds} will inevitably weaken this correlation, adding to the decorrelating effects of thermal smearing. 
This will likely result in significant sensitivities of baryon diffusion coefficients inferred from experimental net baryon rapidity distributions to poorly controlled model ambiguities in the initial space-time rapidity profile of the net baryon density assumed in the dynamical model.   

The baryon diffusion flows observed in the calculations presented in this paper are characterized by an important feature: They show almost no sensitivity to critical effects even for cells passing close to the critical point. Taken at face value, this implies that the hydrodynamic evolution of baryon diffusion leading to the finally emitted ensemble-averaged single-particle momentum spectra does not carry useful information for locating the QCD critical point.

The absence of critical effects on baryon diffusion in this work contrasts starkly with the strong critical effects on the evolution of the bulk viscous pressure found in Ref.~\cite{Monnai:2016kud} which led to significant distortions of the rapidity distributions for all emitted hadron species. The main reason for that behavior is that the bulk viscosity and bulk viscous pressure are generically large around the quark-hadron phase transition, even without explicitly including critical scaling effects. Adding the latter thus causes significant modifications of the dynamics of the bulk viscous pressure.\footnote{%
    We note, however, that the Israel-Stewart equation used in Ref.~\cite{Monnai:2016kud} overestimates the critical effects on bulk viscosity at small frequencies, similar to Eq.~\eqref{eq:Dphi}.}
Baryon diffusion effects, on the other hand, here appear to be insensitivity to critical dynamics, for two main reasons: (i) The baryon diffusion flows are strong at early times but decay very quickly, before the system enters the critical region, because diffusion reduces the initially strong chemical gradients $\nabla(\mu/T)$ that drive the flows, and the baryon diffusion coefficient describing the response to these gradients decreases quickly as the fireball cools by expansion. (ii) By the time the system reaches the phase transition, possibly passing close to the critical point, the Navier-Stokes value of the baryon diffusion current is already very small; critical enhancement by the baryon diffusion coefficient $\kappa_n$ by a power of the correlation length $\xi/\xi_0$ therefore does not help to revive it, and in any case the relaxation rate controlling the approach of the baryon diffusion current to its critically affected Navier-Stokes value is reduced by critical slowing down. The smallness of the baryon diffusion current on the freeze-out surface also implies very small dissipative corrections to the Cooper-Frye formula at particlization.

The observed insignificance of critical effects on baryon diffusion might be taken as permission to calibrate the fireball medium's bulk evolution at BES energies without worrying about the QCD critical point and its location. This may be premature, however, for multiple reasons: (1) The inclusion of shear and bulk viscous effects will modify the expansion trajectories through the phase diagram. Although the critical enhancement of the shear viscosity is negligible ($\eta\sim\xi^{\epsilon/19}$ \cite{RevModPhys.49.435} where $\epsilon{\,=\,}4{-}d$, with $d$ being the number of spatial dimensions), critical slowing down of the  shear stress tensor may still be significant, possibly causing larger residual shear stresses in the vicinity of the critical point than predicted in the absence of critical effects. This has not been studied. The bulk viscous pressure is already known to be strongly affected by critical phenomena \cite{Monnai:2016kud}. We see a potential for these effects to spill over into the baryon diffusion channel through second-order transport coefficients that couple these channels. Should this happen, shear and bulk viscous effects could invalidate some of the findings of the work presented here (in particular the rapid decay of the baryon diffusion current observed in Sec.~\ref{sec5d1}). To clarify this, a future full simulation should include all dissipative effects simultaneously. (2) Large uncertainties still exist about the size of the critical region which is determined by the parametrization of the correlation length \cite{Parotto:2018pwx, Stafford:2021wik}. (3) At lower beam energies, the system may enter the critical region earlier, before the baryon diffusion significantly decays. When the evolution is fully (3+1)-dimensional the edge of the fireball may enter the critical region earlier as well, while the diffusion current is still appreciable.

A fully quantitative evaluation of the significance of critical effects on bulk medium evolution at BES energies can only be made on top of an at least tentatively constrained bulk evolution that includes all necessary theoretical ingredients and complications \cite{Shen:2020jwv}. Only if this confirms negligible critical effects on bulk evolution can the final model calibration be safely made without worrying about critical modifications.

\section*{Acknowledgements}
%
The authors thank Paolo Parotto, Maneesha Pradeep, Krishna Rajagopal, Chun Shen, Mikhail Stephanov and Yi Yin for helpful and stimulating discussions both within and outside the BEST Collaboration. L.D. gratefully acknowledges technical help by Mike McNelis on using the {\sc iS3D} code, by Derek Everett on the freeze-out finder, and by Jan Fotakis on code comparison, and thanks Akihiko Monnai for providing tabulated values for the thermal susceptibility $\chi$ for the {\sc neos} equation of state \cite{Monnai:2019hkn}, as well as Romulo Rougemont for sending tabulated values for the baryon diffusion coefficient $\kappa_n$ obtained in Refs.~\cite{Rougemont:2015wca, Rougemont:2015ona}. L.D. and U.H. were supported in part by the U.S. Department of Energy (DOE), Office of Science, Office for Nuclear Physics, under Award No.~\rm{DE-SC0004286} and within the framework of the BEST Collaboration, and in part by the National Science Foundation (NSF) within the framework of the JETSCAPE Collaboration under Award No.~\rm{ACI-1550223}. U.H. also acknowledges support by a Research Prize from the Alexander von Humboldt Foundation. Computing resources were generously provided by the Ohio Supercomputer Center \cite{OhioSupercomputerCenter1987} (Project PAS0254).

\appendix

\section{Causality analysis near the critical point}\label{sec:causality}

In this Appendix we analyze how the causality (and stability) condition is satisfied near the critical point, focusing on the baryon diffusion only. A more complete analysis involving all non-hydrodynamic degrees of freedom (e.g., shear and bulk stress tensor) is plausible. We evoke small perturbations on top of a flat, homogeneous and static background (denoted by ``$\,\bar~\,$"),
\begin{gather}
    \ed=\bar\ed+\delta\ed(t,x)\,,\quad n=\bar n+\delta n(t,x)\,,\nonumber\\
    u^\mu=\bar u^\mu+\delta u^\mu(t,x)\,, \quad n^\mu=\delta n^\mu(t,x)\,,
\end{gather}
where the perturbations are assumed to be dependent on one spatial coordinate (i.e., $x$) only for simplicity. For a particular direction $x$ we linearize Eqs.~\eqref{eq:conservation} and \eqref{eq:IS_nmu} up to first order in gradient as
\begin{subequations}\label{eq:linearized}
\begin{align}
    &D\delta\ed+\bar w\nabla^x\delta u_x=0\,,\\
    &D\delta n+\bar n\nabla^x\delta u_x+\nabla^x\delta n_x=0\,,\\
    &\bar wD\delta u_x-p_\ed\nabla_x\delta\ed-p_n\nabla_x\delta n=0\,,\\
    &(1+\tau_nD)\delta n_x-\kappa_n(\alpha_\ed\nabla_x\delta\ed+\alpha_n\nabla_x\delta n)=0\,,
\end{align}
\end{subequations}
where
\begin{equation}
\begin{gathered}
    p_e=\left(\frac{\partial p}{\partial e}\right)_n, \quad p_n=\left(\frac{\partial p}{\partial n}\right)_e,\\ 
    \alpha_e=\left(\frac{\partial \alpha}{\partial e}\right)_n, \quad \alpha_n=\left(\frac{\partial \alpha}{\partial n}\right)_e.
\end{gathered}
\end{equation}
Introducing the Fourier component of the linearized variables collectively denoted by $\delta\phi=(\delta e, \delta n, \delta u_\mu, \delta n_\mu)^T$, i.e., 
\begin{equation}
    \delta\tilde\phi(\omega,k)=e^{i\omega t-ikx}\delta\phi(t,x)\,
\end{equation}
where $k=k_x=-k^x$, Eq.~\eqref{eq:linearized} can be transformed to
\begin{equation}
    M\delta\tilde\phi=0
\end{equation}
where
\begin{equation}
    M=\begin{pmatrix}
    i\omega & 0       & ik\bar w & 0 \\
    0       & i\omega & ik\bar n & ik \\
    ikp_e   & ikp_n   & i\omega\bar w & 0 \\
    ik\kappa_n\alpha_e &ik\kappa_n\alpha_n & 0 & i\omega\tau_n+1 \\
    \end{pmatrix}.
\end{equation}
The dispersion relations can be obtained by solving the determinant of the characteristic matrix $M$. For simplicity we assume at this moment $p=p(e)$, thus $p_n=0$ and $c_s^2=p_e$, we then find four eigenmodes:
\begin{equation}
    \omega_1^{\pm}=\pm c_s k\,, \quad \omega_2^{\pm}=i\frac{1\pm \sqrt{1-4\tau_nD_pk^2}}{2\tau_n}\,.
\end{equation}
$\omega_1^{\pm}$ are the modes propagating with the speed of sound; $\omega_2^{\pm}$ are the non-hydrodynamic modes that do not vanish at $k\to0$ (i.e., with finite decay time $\tau_n$), and turn to propagate at large $k$ limit, with the maximum group velocity 
\begin{equation}
v_\text{g}^{\text{max}}=\lim_{k\to\infty}\frac{d\text{Re}\omega}{dk}=\pm\sqrt{\frac{D_p}{\tau_n}}.
\end{equation}
In order to satisfy the causality condition $|v_{\text g}^{\text{max}}|\leq1$, near the critical point, one must demand that $\tau_n$ grows at least as fast as $D_p$, which is obviously manifested, since $\tau_n\sim\xi^2\gg D_p\sim\xi^{-1}$.

\section{Estimating the size of the critical region}
\label{sec:Delta_muT}

In this Appendix we estimate the size of critical region characterize by $\Delta\mu$ and $\Delta T$ (cf. Fig.~\ref{fig:phase_diagram}), as an input to Eq.~\eqref{eq:xi(mu,T)}. Following the analysis and notation convention of \cite{Parotto:2018pwx,Pradeep:2019ccv,Mroczek:2020rpm}, the linear mapping from 3-dimensional Ising variables (i.e., reduced Ising temperature $r$ and magnetic field $h$) to the coordinate variables of QCD phase diagram (i.e., temperature $T$ and baryon chemical potential $\mu$) reads
\begin{equation}\label{eq:mapping_variables1}
    \begin{aligned}
    \frac{T-T_c}{T_c}&=w(r\rho s_1+hs_2)\,,\\
    \frac{\mu-\mu_c}{\mu_c}&=w(-r\rho c_1-hc_2)\,,
    \end{aligned}
\end{equation}
where $(w,\rho)$ are scale factors for the Ising variables $r$ and $h$. For notation simplicity we let $s_i=\sin\alpha_i$, $c_i=\cos\alpha_i$, $i=1,2$, where $\alpha_1$ and $\alpha_2$ are the angles relative to the negative $\mu$ axis for the mapped $r$ and $h$ axis respectively, defined in Fig.~\ref{fig:phase_diagram}. From Eq.~\eqref{eq:mapping_variables1} one can also find the inverse mapping relations
\xa{
\begin{equation}\label{eq:mapping_variables2}
    \begin{aligned}
    r&=\frac{(T-T_c)c_2+(\mu-\mu_c)s_2}{\rho w T_cs_{12}}\,,\\
    h&=-\frac{(T-T_c)c_1+(\mu-\mu_c)s_1}{wT_cs_{12}}\,,
    \end{aligned}
\end{equation}
}
where $s_{12}=\sin(\alpha_1-\alpha_2)=\sin\Delta\alpha$. With the map given by Eq.~\eqref{eq:mapping_variables2}, one can relate the leading singular QCD pressure $p_{\rm crit}(\mu,T)$ and the Gibbs free energy in Ising theory $G(r,h)$ up to a proportional coefficient $A$, i.e.,
\begin{equation}\label{eq:mapping_P-G}
    p_{\rm crit}(\mu,T)=-AG(r(\mu,T),h(\mu,T)).
\end{equation}
The coefficient $A$ can not be determined by universality and thus we parametrize it as 
\begin{equation}
    A=aT_c^4
\end{equation}
where $a$ is an unknown parameter to be chosen.

From Eq.~\eqref{eq:mapping_P-G} one can calculate the singular part of susceptibility, $\chi$, which, as our choice of measure for criticality, diverges near the critical point. Along the crossover ($h=0$) line, 
\begin{equation}
\begin{aligned}
    \chi^{\rm sing}_{h=0}&\sim AG_{\mu\mu}(r,0)\sim AG_{hh}(r,0)h_\mu^2\\
    &\sim A r^{\beta(1-\delta)}\left(\frac{s_1}{wT_cs_{12}}\right)^2\\
    &\sim A \left(\frac{\Delta\mu}{w\rho T_cc_1}\right)^{\beta(1-\delta)}\left(\frac{s_1}{wT_cs_{12}}\right)^2,
\end{aligned}
\end{equation}
while along the $h$ axis ($r=0$), 
\begin{equation}
\begin{aligned}
    \chi^{\rm sing}_{r=0}&\sim AG_{\mu\mu}(0,h)\sim AG_{hh}(0,h)h_\mu^2\\
    &\sim A h^{(1-\delta)/\delta}\left(\frac{s_1}{wT_cs_{12}}\right)^2\\
    &\sim A \left(\frac{c_1\Delta T}{w T_cs_{12}}\right)^{(1-\delta)/\delta}\left(\frac{s_1}{wT_cs_{12}}\right)^2,
\end{aligned}
\end{equation}
where the subscript $\mu$ or $h$ should be understood as the derivative with respect to which with the other variable fixed, for instance, $h_\mu=(\partial h/\partial\mu)_T$. The criterion for the critical phenomena being important is given by the singular part of the susceptibility being comparable to its regular part, i.e.,
\xa{
\begin{subequations}\label{eq:Delta_mu_T1}
\begin{align}
    \chi^{\rm sing}_{h=0}&\sim\chi^{\rm reg}\,,\\
    \chi^{\rm sing}_{r=0}&\sim\chi^{\rm reg}\,.
\end{align}
\end{subequations}
}
Provided $\chi^{\rm reg}\sim T_c^2$, we find correspondingly that
\begin{subequations}\label{eq:Delta_mu_T2}
\begin{align}
    \Delta\mu&\sim T_c\rho w c_1\left(\frac{\sqrt{A}s_1}{w T_c^2s_{12}}\right)^{2/\gamma}, \\
    \Delta T&\sim\frac{(wT_cs_{12})^{-\beta(\delta+1)/\gamma}}{c_1}\left(\frac{\sqrt{A}s_1}{T_c}\right)^{2\beta\delta/\gamma},
\end{align}
\end{subequations}
where $\gamma=\beta(\delta-1)=4/3$.

In order to determine the values of $\Delta\mu$ and $\Delta T$, we adopt the following setup choices:
\begin{gather}\label{eq:setup_Delta_muT}
   \Delta\alpha=\alpha_1-\alpha_2=3^\circ\,, \quad  \rho=\frac{1}{2}\,, \quad w=\frac{1}{2}\,, \quad a=1\,,
\end{gather}
which is consistent with the findings from Ref.~\cite{Pradeep:2019ccv} in the small quark mass limit and the causality requirement from Ref.~\cite{Parotto:2018pwx}. Non-universal parameters given by Eq.~\eqref{eq:setup_Delta_muT} and \eqref{eq:non-universal-parameters} together determine the size of the critical region.

\section{Validation of the numerics}
\label{sec:validation}

\subsection{Longitudinal evolution of baryon diffusion}
\label{sec:long_evo_validation}

\begin{figure}[!htbp]
\begin{center}
\hspace{-0.5cm}
\includegraphics[width= 0.50\textwidth]{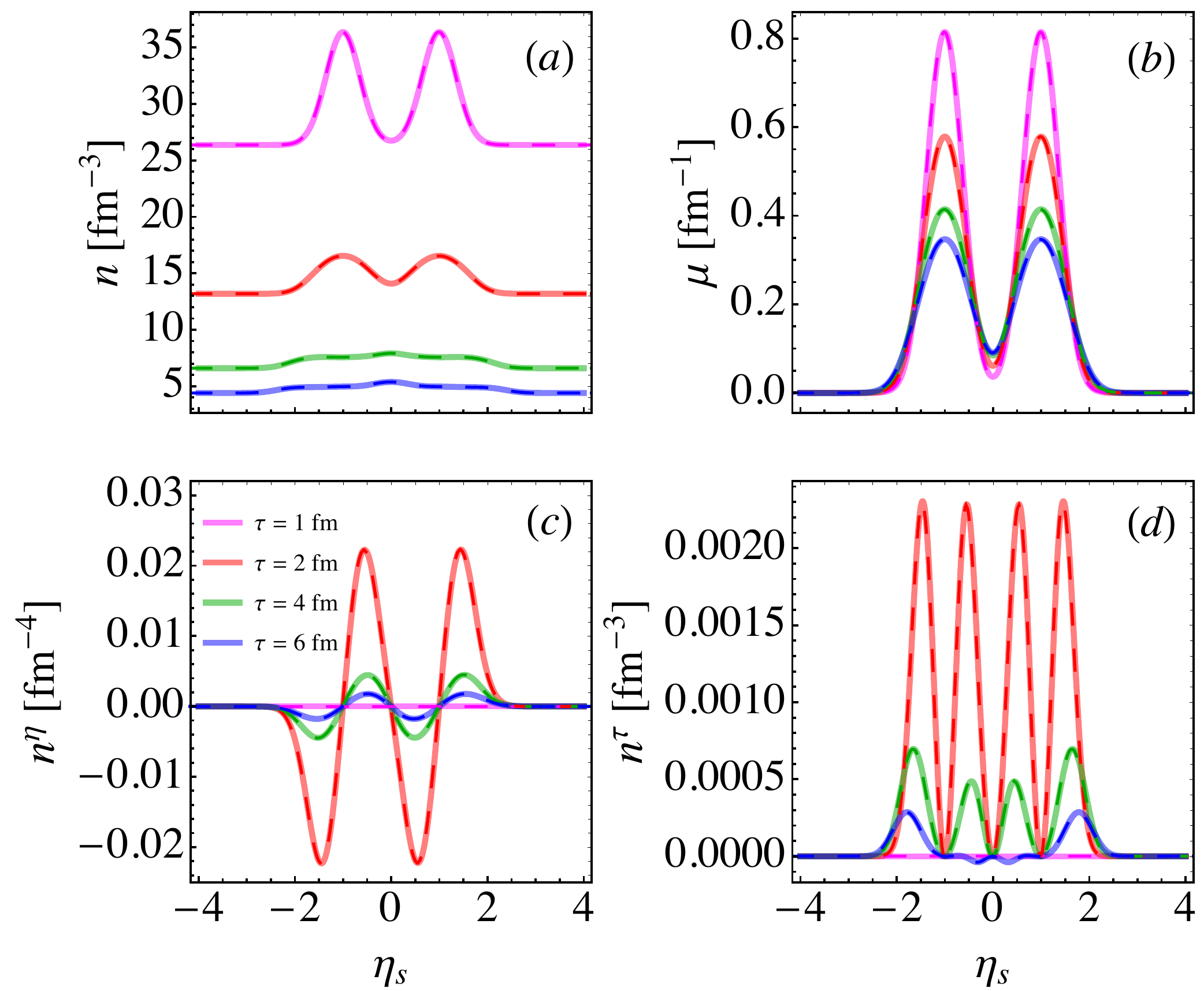}
\caption{%
    Validation of \beshydro{} for solving the hydrodynamic equations \eqref{eq:vhydro-N}-\eqref{eq:simple_eos}, by comparing its numerical results (dashed lines) to semi-analytical solutions (solid lines) at four different times. Here we focus on the baryon sector: (a) baryon density, (b) baryon chemical potential, (c) longitudinal baryon diffusion current and (d) $\tau$-component of baryon diffusion current. The plots demonstrate excellent consistency.}
    \label{fig:difftest}
\end{center}
\end{figure}

\beshydro\ has been validated in different scenarios and found to work at very good precision \cite{Du:2019obx}. In this section, we test its performance in longitudinal dynamics, especially for the baryon diffusion current. For this purpose, in this section we use flow velocity with vanishing transverse components, i.e. $u^\mu=(u^\tau,\,0,\,0,\,u^\eta)$, for a transversely homogeneous system.

Using the decomposition (Eqs.~(\ref{eq:constitutive_ideal1},\ref{eq:constitutive_ideal2})) the conservation laws (Eqs.~(\ref{eq:conservation1},\ref{eq:conservation2})) can be brought into the physically intuitive form \cite{Jeon:2015dfa} 
\begin{eqnarray}
 D\n &=& -\n\theta -\nabla_\mu n^\mu\;,
\label{eq:vhydro-N}
\\
 D\ed &=& -w\theta\;,
\label{eq:vhydro-E}
\\
  w\, Du^\mu &=& \nabla^\mu \peq\;,
\label{eq:vhydro-u}
\end{eqnarray}
where $w=\ed+p$, and the equation of motion for the baryon diffusion is given by
\begin{equation}
    d \V^\mu = \frac{\kappa_n}{\tau_\V} \nabla^{\mu}\alpha - \frac{\V^{\mu}}{\tau_{\V}} - n^\nu u^\mu D u_\nu 
    - u^\alpha \Gamma^\mu_{\alpha\beta} n^\beta\,,
    \label{eq:diff_val}
\end{equation}
in this test. The symbols in the above equations have been introduced in Sec.~\ref{sec:hydro}.

Applying the flow with zero transverse components, these equations can be written into coupled partial differential equations, which can be solved using other softwares (e.g.~{\sc Mathematica}) for testing \beshydro's performance. While these equations clearly exhibit the physics in the LRF (which varies from point to point), \beshydro\ solves the conservation laws (Eqs.~(\ref{eq:conservation1},\ref{eq:conservation2})) in a fixed global computational frame; thus this would be a non-trivial test of our hydrodynamic code.

To solve the equations, we start the system at initial time $\tauI\eq1\,$fm, using the following initial conditions: $u^\mu\eq(1,0,0,0)$ and the initial diffusion current is zero;  the longitudinal distribution of baryon density is given as
\begin{eqnarray}
    n(\tauI, \eta_s) &=& \frac{g_s}{\pi^2}T_0^3 + n_\mathrm{max}\\
    &\times& \left[\exp\left(-\frac{(\eta_s+\eta_+)^2}{\sigma_\eta^2}\right)+\exp\left(-\frac{(\eta_s+\eta_-)^2}{\sigma_\eta^2}\right)\right]\nonumber
\end{eqnarray}
where the initial temperature $T(\tauI, \eta_s)\eq{}T_0\eq0.5\,$GeV, and $n_\mathrm{max}\eq10\,$fm$^{-3}$, $\sigma_\eta\eq0.5$ and $\eta_\pm\eq\pm 1$. Here $g_s\eq16$ is the degeneracy factor in the EoS below. The initial chemical potential is set to be zero everywhere. One also needs the two transport coefficients in Eq. \eqref{eq:diff_val}, and we use \cite{Denicol:2012cn}
\begin{equation}
    \kappa_n = \frac{3}{16\sigma_\mathrm{tot}}\,,\quad \tau_n=\frac{9}{4}\frac{1}{n\sigma_\mathrm{tot}}\,,
\end{equation}
with $\sigma_\mathrm{tot}\eq10\,$mb$\eq1\,$fm$^2$. To close the equations, we use the EoS
\begin{equation}\label{eq:simple_eos}
    e = 3p = 3nT\,,\quad n = \frac{g_s}{\pi^2}T^3\exp{\left(\frac{\mu}{T}\right)}\,.
\end{equation}
We note that the EoS can be analytically inverted to get $T(e, n)$ and $\mu(e, n)$, directly usable for hydrodynamic codes, which is convenient for testing such codes.

\begin{figure}[!tb]
\begin{center}
\hspace{-0.5cm}
\includegraphics[width= 0.50\textwidth]{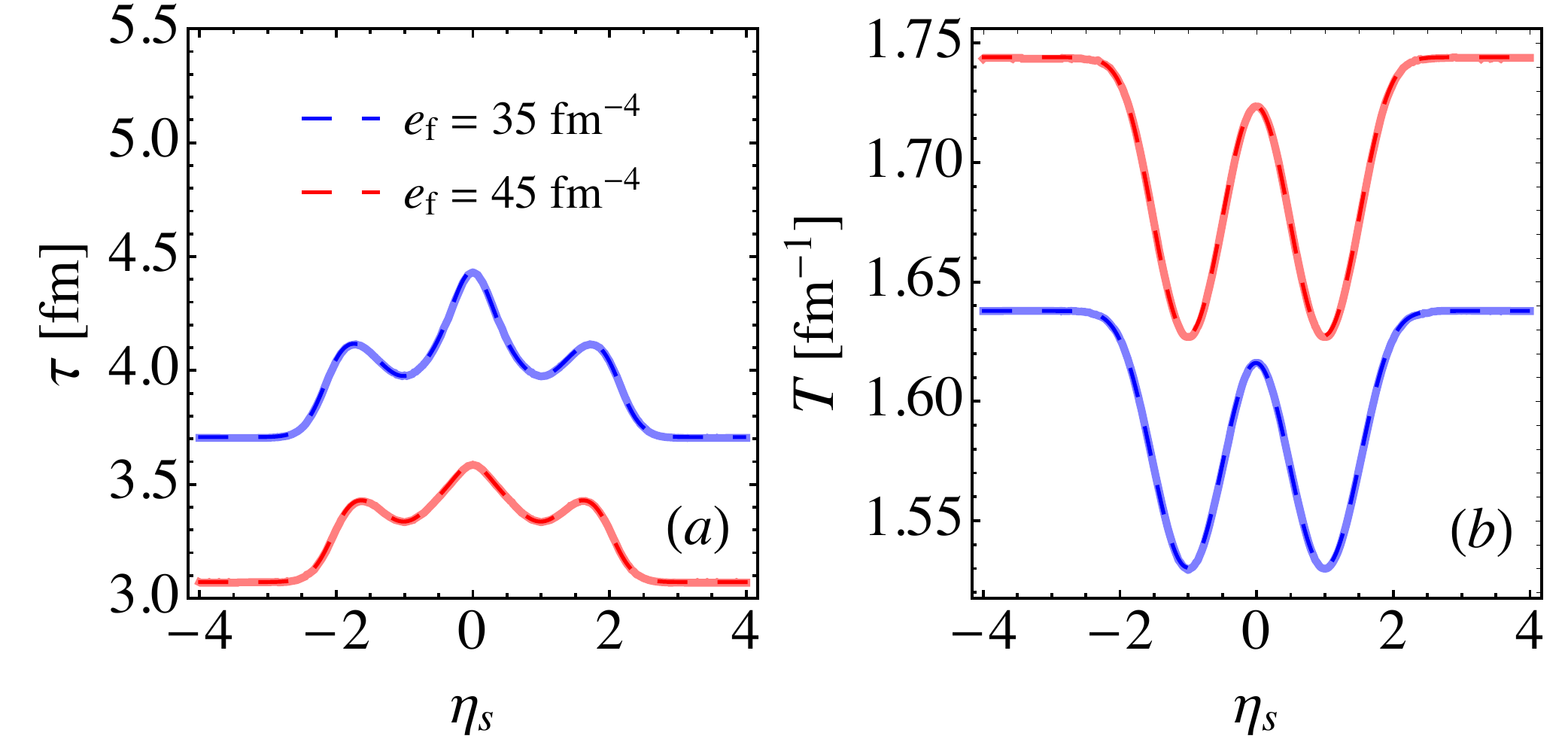}
\caption{%
    Validation of the {\sc Cornelius}-based \cite{Huovinen:2012is} freeze-out finder implemented in \beshydro{}, by comparing the results from \beshydro{} (dashed lines) to semi-analytical results (solid lines), at freeze-out surfaces defined by two different energy densities, $e_\mathrm{f} = 35$\,fm$^{-4}$ (blue lines) and 45\,fm$^{-4}$ (red lines). (a) Space-time profile of the freeze-out surface; (b) corresponding temperature distribution on the surface.}
    \label{fig:fztest}
\end{center}
\end{figure}

The setup above which is from Ref.~\cite{Fotakis:2019nbq} can be solved semi-analytically using e.g. {\sc Mathematica} and used for validating \beshydro, the comparison of which is shown in Fig.~\ref{fig:difftest}, for the baryonic sector. One can see from the figure that \beshydro\ works perfectly well in the longitudinal evolution.

With the solutions, we can also get the space-time distribution of the freeze-out surface, and correspondingly the other hydrodynamic quantities on the surface; thus we can take the chance to test the freeze-out finder implemented in \beshydro{}, which is based on {\sc Cornelius} \cite{Huovinen:2012is}. We have tested its efficiency within \beshydro{} at non-zero baryon density in the transverse plane \cite{du2021jet}. Here, we validate it for this work, where longitudinal dynamics with baryon diffusion current is evolved. In Fig.~\ref{fig:fztest}, we define the freeze-out surface at two different freeze-out energies, and we see that the space-time profile and distribution of the freeze-out temperature perfectly agree with the semi-analytical results.

\subsection{Calculation of various gradients}\label{sec:gradient_validation}

\begin{figure}[!tb]
\begin{center}
    \includegraphics[width= 0.4\textwidth]{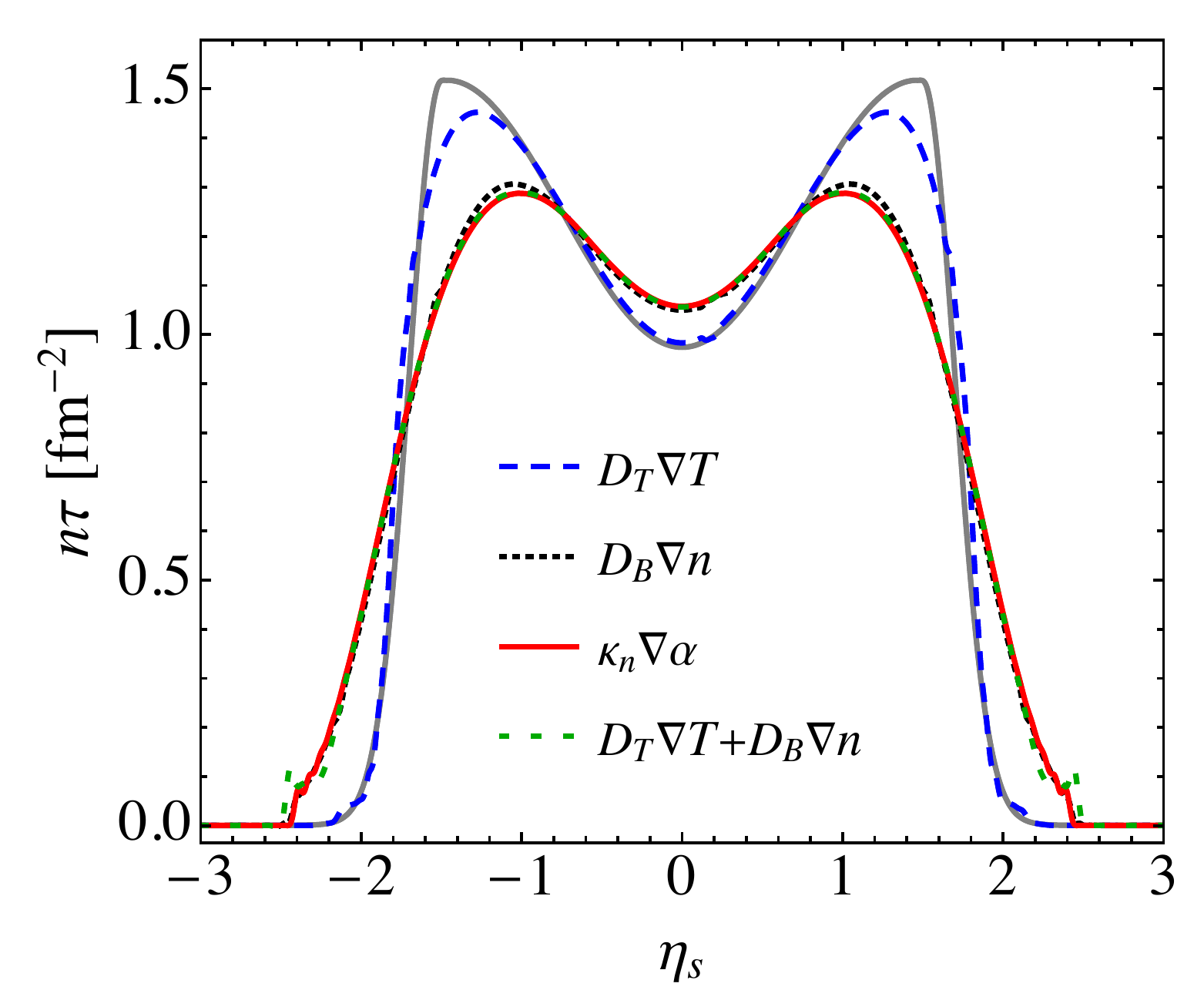}
    \caption{Validation of the equivalence of Eqs.~\eqref{eq:kappa} and \eqref{eq:chib}, and comparisons among contributions from different gradient terms in Eqs.~\eqref{eq:kappa} and \eqref{eq:chib} to baryon diffusion current. Note that consistence between results from $\kappa_n\nabla{}\alpha$ (red solid line) and $D_T\nabla{}T+D_B\nabla{}n$ (green dashed line) is a highly non-trivial test of numeric methods, since the later involves calculating $(\partial p/\partial T)_n$ and interpolating tabulated $\chi$ etc. Though the overall agreement is very good, small wiggles can be seen near the edge for the latter case, a reflection of the complexities in numerics.}
    \label{fig:gradtest}
\end{center}
\end{figure}

As mentioned in Sec.~\ref{sec:hydro}, to include critical contributions to the EoS without using such an EoS, we rewrite the Navier-Stokes limit term of baryon diffusion into two terms: 
\begin{eqnarray}
n^{\mu }_\mathrm{NS} &=& \kappa_n \nabla^{\mu}\alpha\,\label{eq:kappa}\\
&=&\frac{\kappa_n}{T\chi}\nabla^{\mu}n +\frac{\kappa_n}{Tn}\left[\left(\frac{\partial p}{\partial T}\right)_n-\frac{w}{T}\right]\nabla^{\mu}T\,,\label{eq:chib}
\end{eqnarray}
through which critical behavior \eqref{eq:coeff_scaling} is introduced to the thermal properties of the system. In the original \beshydro\, Eq.~\eqref{eq:kappa} is calculated and tested, but apparently Eq.~\eqref{eq:chib} is much more complicated to calculate numerically.

Here we validate the code by verifying that exactly the same results can be achieved using the two expressions for the Navier-Stokes term in Eqs.~(\ref{eq:kappa},\ref{eq:chib}). A quick and easy test can be done using the setup from Sec.~\ref{sec:long_evo_validation}, where
\begin{equation}
    \chi = \left(\frac{\partial^2p}{\partial\mu^2}\right)_T = \frac{n}{T}\,,\quad \left(\frac{\partial p}{\partial T}\right)_n = n\,,\label{eq:chi_dpdt}
\end{equation}
are analytically available and easily used in \beshydro. We have checked that both methods in Eqs.~(\ref{eq:kappa},\ref{eq:chib}) give precisely the same results in Fig.~\ref{fig:difftest}. However, in the simulations of this work, the {\sc neos} is used, and the two terms in Eq.~\eqref{eq:chi_dpdt} are not analytical calculable. Thus we repeat the comparison to MUSIC as in Ref.~\cite{Du:2019obx} for the two gradients, and show the validation in Fig.~\ref{fig:gradtest}, which show excellent agreement. From the figure, we also see that the baryon transport driven by $D_B\nabla^{\mu}n$ is very close to that driven by $D_B\nabla^{\mu}n+D_T\nabla^{\mu}T$, indicating that density gradients dominates over temperature gradients in the Navier-Stokes limit of the baryon diffusion current.


\bibliography{references}

\end{document}